\documentclass[12pt,a4paper]{article}
\usepackage{epsfig}
\usepackage[latin1] {inputenc}
\usepackage{amsmath} \usepackage{amsfonts} \usepackage{amssymb} \usepackage{graphicx} \usepackage{longtable}
\author{Les Hatton \\ leshatton.org\thanks{lesh@oakcomp.co.uk}}
\title{Power-laws and the Conservation of Information in discrete token systems:\\Part 1 General Theory}
\begin{document}
\maketitle
\bibliographystyle{plain}
\begin{abstract}
The Conservation of Energy plays a pivotal part in the development of the physical sciences.  With the growth of computation and the study of other discrete token based systems such as the genome, it is useful to ask if there are conservation principles which apply to such systems and what kind of functional behaviour they imply for such systems. 

Here I propose that the \textit{Conservation of Hartley-Shannon Information plays the same over-arching role in discrete token based systems as the Conservation of Energy does in physical systems}.  I will go on to prove that this implies power-law behaviour in component sizes in software systems no matter what they do or how they were built, and also implies the constancy of average gene length in biological systems as reported for example by \cite{XuJune2006}.

These propositions are supported by very large amounts of experimental data extending the first presentation of these ideas in \cite{Hatton2011a}.
\end{abstract}
\begin{verbatim}
Keywords: Information content, gene length,
Component size distribution, Power-law
\end{verbatim}
\section{Preliminaries}
\subsection{Conservation of Energy}
The Conservation of Energy is one of a few principles which are at the very heart of all physical systems.  The principle has been modified over the years notably to take account of the 4-vectors of relativity and mass-equivalence but it remains pivotal.  In the same year as Einstein's eponymous paper on general relativity (1915), Emmy Noether proved a remarkable theorem which amongst many other things shows that the principle of conservation of energy is a consequence of general invariance under time translations.

The study of discrete systems is much younger and has only come of age in the digital age where we now routinely write millions of lines of source code to analyse terabytes of digital data.  It is of great interest to see if there are similarly fundamental principles which apply to the evolution of discrete systems.

This paper identifies such a principle and produces very large amounts of supporting evidence.  The paper brings together various concepts which will be individually described briefly now.

\subsection{Power-laws}
Power-law behaviour can be represented by the pdf (probability density function) p(s) of entities of a certain size s appearing in some process, being given by a relationship like:-

\begin{equation}
p(s) = \frac{k}{s^\alpha}   \label{eq:power}
\end{equation}

where k is a constant, which on a log p - log s scale is a straight line with negative slope $- \alpha$.  It can easily be verified that the equivalent cdf (cumulative density function) c(s) derived by integrating (\ref{eq:power}), also obeys a power-law, (for $\alpha \neq 1$).  For noisy data, the cdf form is most often used because of its fundamental property of reducing noise, as noted by \cite{Newman2006} and it is this form which will be used here.

Power-law behaviour has been studied in a very wide variety of environments, see for example \cite{Zipf35} (linguistics), \cite{Rawl04} (economic systems) and the recent excellent reviews by \cite{Mitzenmacher2003} and \cite{Newman2006}.  In software systems there has been significant activity, much of it recent, \cite{ClarkGreen77}, \cite{Mitzenmacher2002}, \cite{Myers2003}, \cite{Challet2004}, \cite{Gorshenev2004}, \cite{Potanin05}, \cite{Baxter06},  \cite{Concas2007} and \cite{HatTSE08} all discuss power-law behaviour but in rather different contexts.

Mitzenmacher \cite{Mitzenmacher2002} considers the distributions of file sizes in general filing systems and observed that such file sizes were typically distributed with a lognormal body and a Pareto (i.e. power-law) tail.

In comparison, Gorshenev and Pis'mak \cite{Gorshenev2004} studied the version control records of a number of open source systems with particular reference to the number of lines added and deleted at each revision cycle.

In this paper, I will step back from this and look for more fundamental reasons why power-laws are so ubiquitous.

\subsection{Systems of discrete choices}

A system based on discrete choices is any system which is built from discrete pieces based on some available set of choices.  The genome is a perfect example.  This is an exceptionally complex system which has evolved over hundreds of millions of years, astonishingly from a set of only four choices, the four bases of the genetic alphabet, adenine, cytosine, guanine and thiamine, (acgt).  The human genome comprises some 3 billion such bases.  I will refer to such choices as an \textit{alphabet}.

As an example, the first 60 bases of the measles virus\footnote{http://www.ncbi.nlm.nih.gov/nuccore/HM562900.1} are

\begin{verbatim}
atggactcgc tatctgtcaa ccagatcttg taccccgaag ttcacctaga tagcccgata
\end{verbatim}

Computational science provides many more examples.  In computational science, the source code of every computer program written by every researcher in pursuit of their computational results uses one or more programming languages.  Such programs form an essential part of the vast majority of modern scientific work and parenthetically present a huge challenge to scientific reproducibility\cite{Ince2012}.

The individual bases or alphabet of a programming language are called \textit{tokens} and may take two forms; the \textit{fixed} tokens of the language as provided by the language designers and the \textit{variable} tokens.  Fixed tokens include (in the languages C and C++ for example) \textit{if}, \textit{else}, \textit{while}, \textit{\{}, \textit{\}}.  These can not be changed, the programmer can only choose to use them or not. Variable tokens, with some small lexical restrictions, can be arbitrarily invented by the scientific programmer whilst constructing their program.  These might be identifier names such as \textit{numberOfCandidateCollisions} or \textit{lengthOfGene} or constants such as 3.14159265.  Computational scientific systems and indeed every other form of software system evolve from such tokens.  There are many programming languages but all obey the same principles.

Such programs are often very large.  The software deployed in the search for the recently discovered Higg's boson comprises around 4 million lines of code \cite{Rousseau2012}.  At an average of around 6 tokens per line of code, this corresponds to some 20-25 million tokens, although this is still less than 1\% of the human genome.  The largest systems in use today appear to be around 100 million lines of source code\cite{Mossinger2010}, perhaps 15\% of the number of tokens of the human genome.

As an example, consider the following simple\footnote{Simple in the sense that nobody ever uses it because far faster sorting mechanicsms are known but it is useful for teaching purposes.} bubblesort algorithm written in C, for example \cite{Sedgewick1990}.

\begin{verbatim}
void bubble( int a[], int N)
{
  int i, j, t;
  for( i = N; i >= 1; i--)
  {
    for( j = 2; j <= i; j++)
    {
      if ( a[j-1] > a[j] )
      {
        t = a[j-1]; a[j-1] = a[j]; a[j] = t;
      }
    }
  }
}
\end{verbatim} 

This algorithm contains 94 tokens in all based on 18 of the fixed tokens of ISO C

\begin{verbatim}
void int ( ) [ ] { , ; for = >= -- <= ++ if > -
\end{verbatim}

and the 8 variable tokens (i.e. invented by the programmer)

\begin{verbatim}
bubble a N i j t 1 2
\end{verbatim}

Although programming languages have a much richer alphabet of tokens than genes, they obey the same principles - some external process chooses tokens from the available alphabet.  I will argue that this process is driven by a beautiful underlying clockwork, that of Conservation of Information.

\subsection{Information theory}

\begin{figure}
\includegraphics[width=12cm]{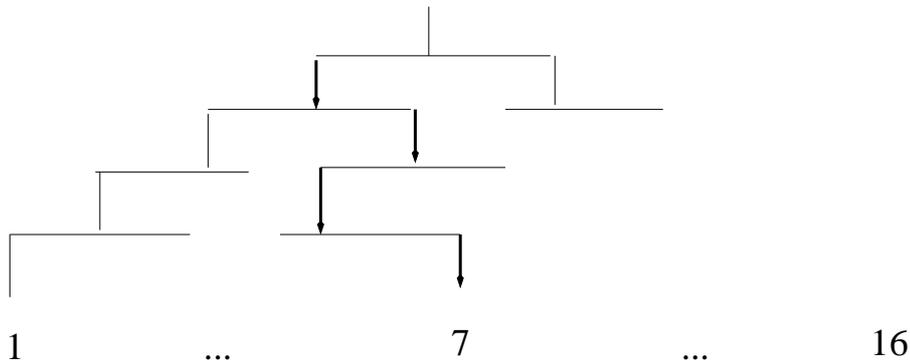}
\caption{A binary tree.  Each level proceeding down can either go left or right.  There are four levels leading down to one of $2^{4}$ = 16 possibilities.  Only 4 choices are needed to reach any of the possibilities.  We note that $log_{2}(16) = 4$.  Here the number 7 has been singled out by the choices left, right, left, right.}
\label{fig:choices}
\end{figure}

Information theory has its roots in the work of Hartley\cite{Hartley1928} who showed that a message of N signs (i.e. tokens) chosen from an \textit{alphabet} or code book of S signs has $S^{N}$ possibilities and that the \textit{quantity of information} is most reasonably defined as the \textit{logarithm} of the number of possibilities or choices.  To gain a little insight into the reason why the logarithm makes sense, consider Figure \ref{fig:choices}.  The number of choices necessary to reach any of the 16 possible targets is the number of levels which is the $log_{2}$(number of possibilities).  The base of the logarithm is not important here.

Information theory was developed very substantially by the pioneering work of Shannon\cite{Shannon1948}, \cite{Shannon1949}.  However it is important not to conflate information content with functionality or meaning and Cherry\cite{Cherry1963} specifically cautions against this noting that the concept of information based on alphabets as extended by Shannon and Wiener amongst others, \textit{only relates to the symbols themselves} and not their \textit{meaning}.  Indeed, Hartley in his original work, defined \textit{information} as the successive selection of signs, rejecting all meaning as a mere subjective factor. In the sense used here therefore, Conservation of Information will be synonymous with Conservation of Choice, not meaning.  This turns out to be enough to predict important system properties.  In other words, those properties depend only on the alphabet and not on what combining tokens of the alphabet might mean in any human sense.

\section{A statistical mechanical model of a discrete tokenised system}
Armed with these pieces of information, I will now describe a variational model in which Conservation of Information is a fundamental constraint as first described by \cite{Hatton2011a}.  A general discrete tokenised system will be considered here as T tokens distributed in some way amongst M non-nested \textit{components}, each containing $t_{i}$ tokens where i = 1, ... ,M.  In software systems, a component might be a subroutine (Fortran), function (C) or procedure (Tcl-Tk).  In OO languages, it might be a method\footnote{Strictly speaking methods can be and usually are nested in OO systems although when compiled they are simply treated as a function with some context and so remain relevant to this model of M non-nested components.  Proper source analysis requires them to be treated the same way.}.  In a genetic system, a component might be a gene.

Then the number of ways of organising this system is given by:-

\begin{equation}
W=\frac{T!}{t_{1}!t_{2}!..t_{M}!} \label{eq:comb}
\end{equation}

where

\begin{equation}
T = \sum_{i=1}^{M} t_{i}  \label{eq:con1}
\end{equation}

Also suppose there is some externally imposed entity $\varepsilon_{i}$ associated with each token of component \textit{i} whose total amount is given by

\begin{equation}
U = \sum_{i=1}^{M} t_{i} \varepsilon_{i} \label{eq:con2}
\end{equation}

Using the method of Lagrangian multipliers as described in \cite{HatTSE08}, the most likely distribution satisfying equation (\ref{eq:comb}) subject to the constraints in equations (\ref{eq:con1}) and (\ref{eq:con2}) will be found.  This is equivalent to maximising the following variational derived by taking the log of (\ref{eq:comb}).

\begin{equation}
log W = T log T - \sum_{i=1}^{M} t_{i} log (t_{i}) + \lambda \lbrace T - \sum_{i=1}^{M}t_{i} \rbrace + \beta \lbrace U -  \sum_{i=1}^{M} t_{i} \varepsilon_{i} \rbrace    \label{eq:min1}
\end{equation}

where $\lambda$ and $\beta$ are the multipliers and log is the natural logarithm.  Setting $\delta (log W) = 0 $ and using the assumption that T and the $t_{i}$ are both $\gg 1$ leads to

\begin{equation}
0 = - \sum_{i=1}^{M} \delta t_{i} \lbrace log (t_{i}) + \alpha + \beta \varepsilon_{i} \rbrace    \label{eq:var1}
\end{equation}

where $ \alpha = 1 + \lambda $.  This must be true for all variations $\delta t_{i} $ and so

\begin{equation}
log (t_{i}) = - \alpha - \beta \varepsilon_{i}    \label{eq:var2}
\end{equation}

Using equation (\ref{eq:con1}) to replace $\alpha$, this can be manipulated into the most likely, i.e. the equilibrium distribution

\begin{equation}
t_{i} = \frac{T e^{-\beta \varepsilon_{i}}}{\sum_{i=1}^{M} e^{-\beta \varepsilon_{i}}}    \label{eq:varn}
\end{equation}

Defining $p_{i} = \frac{t_{i}}{T}$, equation (\ref{eq:varn}) then yields

\begin{equation}
p_{i} = \frac{e^{-\beta \varepsilon_{i}}}{\sum_{i=1}^{M} e^{-\beta \varepsilon_{i}}}    \label{eq:varp}
\end{equation}

Following \cite{Rawl04} and  and referring to equation (\ref{eq:con2}), $p_{i}$ can then be interpreted as the probability that a component of size $t_{i}$ tokens is found is exponentially related to $\varepsilon_{i}$.  The larger $\varepsilon_{i}$ for example, the less likely such a component is to appear.

Hinting at what is to come, \textit{we can see immediately that in any discrete system where $\varepsilon_{i}$ is the logarithm of some quantity, then the resulting size distribution is overwhelmingly likely to be power-law since exp( -c log d ) = $d^{c}$.}

\subsection{Merging with information theory}
I will now repeat the argument I gave in \cite{Hatton2011a}.  Suppose then that the \textit{unique} alphabet of the $i^{th}$ component contains $a_{i}$ tokens and as defined above, $t_{i}$ tokens in all.  The number of ways of arranging the tokens of this alphabet in component \textit{i} is therefore $a_{i}^{t_{i}}$. Following Hartley, the quantity of information in component \textit{i} will therefore be defined as

\begin{equation}
 I_{i} = log(a_{i})^{t_{i}}= t_{i} log a_{i}  \label{eq:hfunc}
\end{equation}

To blend this into the variational method shown earlier, the same computational device used in \cite{HatTSE08} was used.  I will introduce I, the total amount of information, as the sum of the information in each component as follows:-

\begin{equation}
I = \sum_{i=1}^{M} t_{i} (\frac{I_{i}}{t_{i}}) \label{eq:con4}
\end{equation}

I will now identify $\varepsilon_{i}$ with $(\frac{I_{i}}{t_{i}})$ in equation (\ref{eq:varn}).  In other words, each token of component \textit{i} has an externally imposed information density associated with it given by $(\frac{I_{i}}{t_{i}})$.  This assumes that the information per token in a single component takes some average value but that this can vary amongst components.  This seems reasonable in that it suggests that no particular token is any more important than any other when developing a particular component as some functional entity, however it allows for the fact that this can vary amongst different components which fits well with intuition that some functional entities are in some sense more complicated than others.  Note finally that introducing this additional functional dependence of $\varepsilon_{i}$ on $t_{i}$ does not disrupt the development which led to equation (\ref{eq:varp}) as $\varepsilon_{i}$ is fixed externally by assumption.

Equation (\ref{eq:varp}) can then be written as

\begin{equation}
p_{i} = \frac{e^{-\beta \frac{I_{i}}{t_{i}}}}{Q(\beta)}    \label{eq:varf}
\end{equation}

where

\begin{equation}
Q(\beta) = \sum_{i=1}^{M} e^{-\beta \frac{I_{i}}{t_{i}}}
\end{equation}

Combining equations (\ref{eq:varf}) and (\ref{eq:hfunc}) then gives

\begin{equation}
p_{i} = \frac{e^{- \beta log a_{i}}}{Q(\beta)}  \label{eq:logs}
\end{equation}

So we finish up with the following predicted power-law distribution

\begin{equation}
p_{i} = \frac{(a_{i})^{- \beta}}{Q(\beta)} \sim (a_{i})^{- \beta}  \label{eq:pwrlaw}
\end{equation}

subject to the twin constraints that the total number of tokens T is fixed

\begin{equation}
T = \sum_{i=1}^{M} t_{i}         \label{eq:T}
\end{equation} 

and the total Hartley / Shannon information content I is also fixed

\begin{equation}
I = \sum_{i=1}^{M} I_{i}	\label{eq:I}
\end{equation}

where $I_{i}$ is the information content of the $i^{th}$ component

It is worth repeating that this overall process \textit{does not care about the tokens themselves} - all individual microstates are equally likely.  It simply says that if total size and choice in the Hartley-Shannon sense are conserved during the process of distributing the tokens, then power-law distribution of component size in the unique alphabet of tokens used is overwhelmingly likely to emerge since it occupies the vast majority of the microstates.

Given its central position in what follows, it is useful to retrace the assumptions.  These are
\begin{enumerate}
\item The variational method assumes that both $t_{i}$ and T are $\gg 1$.  This turns out to be a very good approximation for nearly all the data here.  Components with $t_{i} < 10$ are rare indeed in software systems because of the token overhead described earlier and genes are typically much longer.
\item The variational method \textit{enforces that the total size T is kept constant} whilst the most likely solution is found.  It should be noted that this is not its actual size at any point in time, but the eventual size defined by its intended functionality in an ergodic sense.  In other words, if the same system was produced many times independently, then for a particular T, the variational method finds the most likely distribution of $t_{i}$ subject to the constraints.
\item The variational method also \textit{enforces the Conservation of Information}.  I is \textit{not} the same as functionality, it is simply related to choice from the available alphabet in the Hartley-Shannon sense.
\end{enumerate}

\section{Application to software systems}

\subsection{Software components and tokens}
Following the earlier discussion, we can write the unique alphabet of $a_{i}$ tokens in the $i^{th}$ component of a software system of M components as

\begin{equation}
a_{i} = a_{f} + a_{v}(i)     \label{eq:toks}
\end{equation}

where $a_{f}$ is the alphabet of fixed tokens and $a_{v}(i)$ is the alphabet of invented tokens and is clearly dependent on i, since programmers are free to create them as and when desired.

It will be noted that $a_{f}$ is taken as independent of component whereas $a_{v}(i)$ is dependent on the component.  To flesh this out a little, it is worthwhile introducing a highly relevant property of programming languages at this point.  \textit{Smaller components tend to be dominated by tokens fixed by the programming language and larger components tend to be dominated by tokens invented by the programmer, for example constants and identifier names.}  The reasons for this are first, the fixed tokens of a language are limited in number and a significant number of these are very rarely used, (for example, the 10 \textit{trigraphs} or the \textit{goto} in ISO C). Second, there is a certain token overhead which must be paid in order to produce the simplest of components.  As the component size grows, the fixed token alphabet rapidly stabilises whilst the invented token alphabet grows without any such limits.  It is therefore a reasonable assumption to consider the alphabet of fixed tokens as approximately constant across components.

To support this conclusion, throughout these studies the (variable/fixed) token ratio was found to be typically 0.3 or less for the small components.  In contrast, the same token ratio is typically greater than 5 for large components.  In addition, on average, the fixed token population does not vary with component size - linear regression of $a_{f}$ against $t_{i}$ on the 526,158 components extracted in this study revealed a gradient of around $7.0 \times 10^{-4}$, in other words it is effectively zero.

In other words, as the component size grows, the fixed token alphabet hardly changes in this dataset whilst the variable token alphabet grows without any such limits.  For example, more than 95\% of the components analysed here used less than 30 fixed tokens.

This has a profound effect on the predicted shape of the distribution as we will see.

\subsection{Predicted shape of the size distribution}
In anticipation of applying this to software data, it is conventional to consider the cdf (cumulative distribution function) as discussed earlier, rather than the pdf because of its much more stable behaviour in the presence of noise \cite{Newman2006}.  This is given by integration of (\ref{eq:pwrlaw}) as

\begin{equation}
c_{i} \sim a_{i}^{-\beta+1}      \label{eq:cdf1}
\end{equation}

for $\beta \ne 1$.  It is then possible to anticipate the approximate predicted shape of the size distribution as follows.  For small components, we have seen that it is reasonable to assume that the number of fixed tokens will tend to dominate the total number of tokens because of the fixed token overhead.  In other words, $a_{f} \gg a_{v}(i)$.  Equation (\ref{eq:cdf1}) can then be written

\begin{equation}
c_{i} \sim (a_{f})^{- \beta+1} (1 + \frac{a_{v}(i)}{a_{f}})^{- \beta+1}    \label{eq:cdf2}
\end{equation}

In other words,

\begin{equation}
c_{i} \sim (a_{f})^{- \beta+1}   \label{eq:cdf3}
\end{equation} 

which implies that $c_{i}$ \textit{will be tend to a constant for small components on a log-log plot}.  For large components, using the same arguments, the general rule applies

\begin{equation}
c_{i} \sim (a_{i})^{- \beta+1}   \label{eq:cdf4}
\end{equation}

\begin{figure}
\includegraphics[width=8cm,height=6cm]{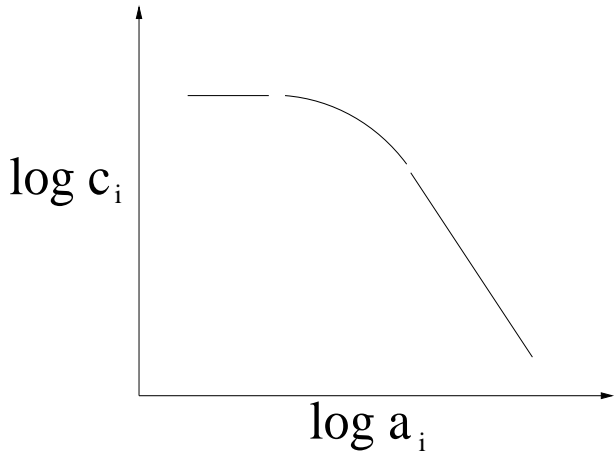}
\caption{The predicted cdf using the model described in this paper.  The cdf is predicted to be approximately constant for small components and power-law in $a_{i}$ for large ones with a merging zone between.}
\label{fig:prediction}
\end{figure}

The generic shape of the resulting predicted curve on a log-log scale is shown in Figure \ref{fig:prediction}.

\subsection{Experimental verification}
An unusually large number of systems were analysed across multiple languages in order to increase the statistical relevance.  Open source has had many benefits but one of particular value to researchers is the enormous amount of source code which can be freely downloaded, often with excellent development history.  In this study, 6 languages were chosen, Ada, C, C++, Fortran, Java and Tcl.  This covered a very wide variety of implementation areas and paradigms.  In these languages, around 90 packages were downloaded comprising some 55.5 million lines of code over many development areas and almost half a billion tokens in all.  These include for example, the whole of the Linux kernel, PHP, X11R7, Postgresql and Perl (C), the Ada validation system (Ada), the KDE desktop (C++) and the Java Virtual machine (Java).  As well as these, the author had access to some commercial systems in Tcl, Fortran and C but these only totalled around 2\% of the total code analysed.  Individual package sizes spanned 3 orders of magnitude varying from around 10 KSLOC (thousands of source lines of code) to some 13.5 MSLOC (millions of source lines of code) as is found in version 2.6.27 of the Linux kernel.

\subsubsection{Lexical analysis}
The extraction of tokens from a program is not a trivial process and requires the development of tools which mimic the front-end of compilers \cite{Aho1977}.  The minimum requirements for a lexical analyser for each language considered here, were

\begin{itemize}
\item The ability to extract tokens and to distinguish between the two token forms, fixed and variable.
\item The ability to recognise the start and end of a component.  This is simpler for non-OO languages than OO languages because the latter admit nested components or methods.  In this analysis a useful approximation is that nested methods can be ignored.  This has a small effect as noted later.
\end{itemize}

The resulting generic tokeniser was written in C for optimal performance and also to exploit the well-known \textit{lex} tool for generating lexical analysers\footnote{, http://flex.sourceforge.net/manual/}.  It comprises around 2000 lines of C and 1300 lines of lex\footnote{For the purposes of open verification, it is included at http://www.leshatton.org/category/scientific-writing/datasets/.}.  There may be certain difficult tokens in some languages which are simply ignored by this generic analyser.  This excluded only a tiny fraction of components from the analysis however.  As a quality control check, the C and Fortran analysers were checked against and found to agree closely with existing full parsers written by the author some years ago, both of which parsed the relevant compiler validation suites correctly, (FIPS160 in the case of ISO C90 and the ACVS in case of Fortran 77).  The resulting generic tokeniser is extremely fast and can extract tokens at around the rate of 100,000 lines a second on a typical Linux desktop allowing the analysis of the very large amounts of source code considered here.

\subsubsection{Results}
\begin{figure}
\includegraphics[width=8cm,height=6cm]{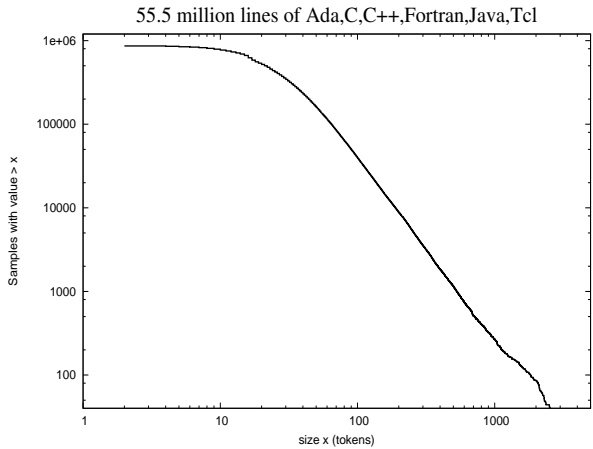}
\caption{The measured cdf for all 90 systems analysed here combined into one super-system.  This comprises around 13\% Java, 6\% C++, 8\% Fortran, Ada and Tcl combined and around 73\% C.  This very roughly reflects the amount of each language freely available on open source, although 24 million lines came from Linux and BSD which are in C.  Absenting these unusually large systems, the percentages are 22\%, 10\%, 15\%, 53\% respectively.}
\label{fig:universe}
\end{figure}

Using the generic lexical analyser described in the previous section, All 90 or so systems were analysed comprising some 55.5 million lines of code in six languages.  To emphasise that the nature of the tokens or their meaning does not really matter ergodically, Figure \ref{fig:universe} shows the measured cdf for the whole dataset together comprising almost half a billion tokens.

This can be compared with the prediction represented by Figure \ref{fig:prediction}.  The two are emphatically alike.  The predicted linearity of the power-law tail of Figure \ref{fig:universe} was subjected to a standard test for significance using the linear modelling function lm() in the widely-used R statistical package\footnote{http://www.r-project.org/}.  This reported a very high degree of linearity with a linear-fit correlation of 0.998 between token counts of 30 and 3000, a span of two decades.  The same analysis reports a slope of -2.125 +/ 0.003, which is in the range -2 $\rightarrow$ -3 reported for most natural phenomena by \cite{Newman2006}.  The associated p-value, (the probability of finding a dataset more unlikely than this one by chance) is $< 2.2e-16$, an extremely emphatic result.  The corresponding output from R is shown below.

\begin{verbatim}
> source("plot_tail.R")
Read 957 items
Read 957 items
> summary(fm)

Call:
lm(formula = y ~ x, data = universe)

Residuals:
     Min       1Q   Median       3Q      Max 
-0.25510 -0.05790 -0.02892  0.05866  0.33697 

Coefficients:
             Estimate Std. Error t value Pr(>|t|)    
(Intercept) 20.285185   0.019336  1049.1   <2e-16 ***
x           -2.125199   0.003133  -678.4   <2e-16 ***
...

Residual standard error: 0.08646 on 955 degrees of freedom
Multiple R-squared: 0.9979,     Adjusted R-squared: 0.9979 
F-statistic: 4.603e+05 on 1 and 955 DF,  p-value: < 2.2e-16 
\end{verbatim} 

The qualitative prediction of the asymptotic constant behaviour of the cdf for small components is also reassuring.  It can be concluded that this experiment very strongly supports the model presented in \cite{Hatton2011a} and repeated here as (\ref{eq:pwrlaw}).  The resulting behaviour implicit in Figure \ref{fig:universe} contrasts nicely with the pure straight line predicted for monkeys pounding on keyboards as eloquently described by \cite{Mitzenmacher2003}.  The ergodic nature of (\ref{eq:pwrlaw}) simply accumulates all possible programmers pounding on keyboards.  As will be seen, it also works well with much smaller numbers, i.e. individual systems, a characteristic of classical statistical mechanics.

\subsubsection{Individual systems}
\begin{figure}
\centering
\begin{tabular}{cc}
\epsfig{file=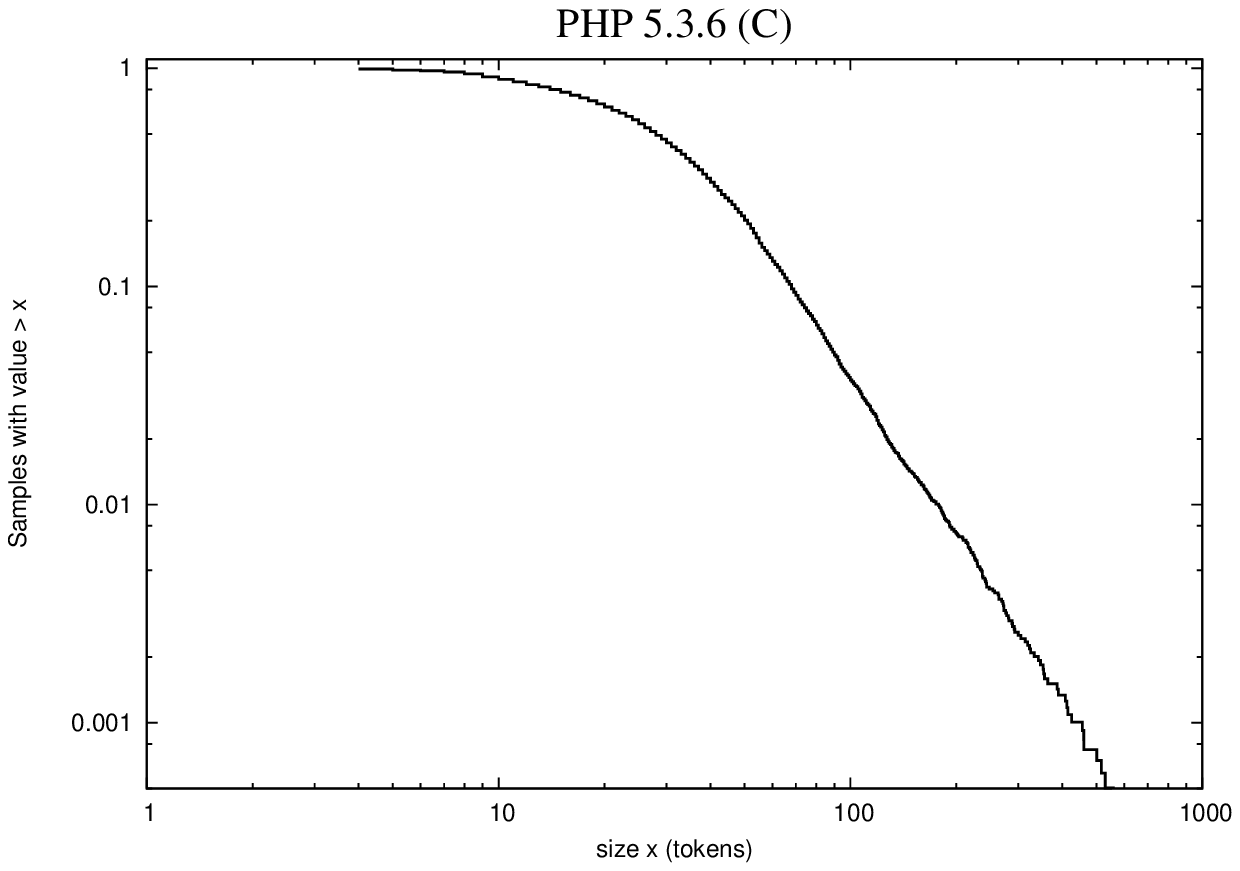,width=0.5\linewidth,clip=} &
\epsfig{file=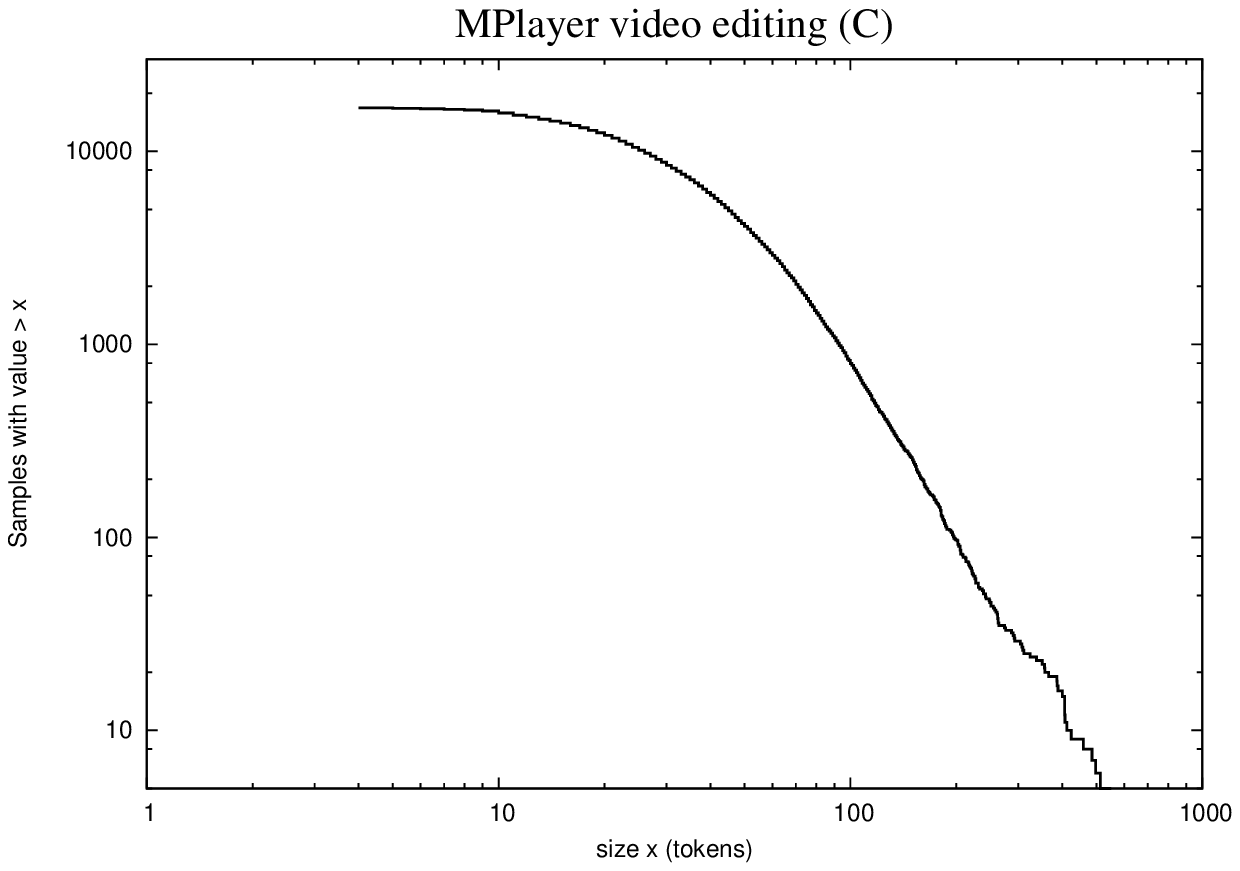,width=0.5\linewidth,clip=} \\
\epsfig{file=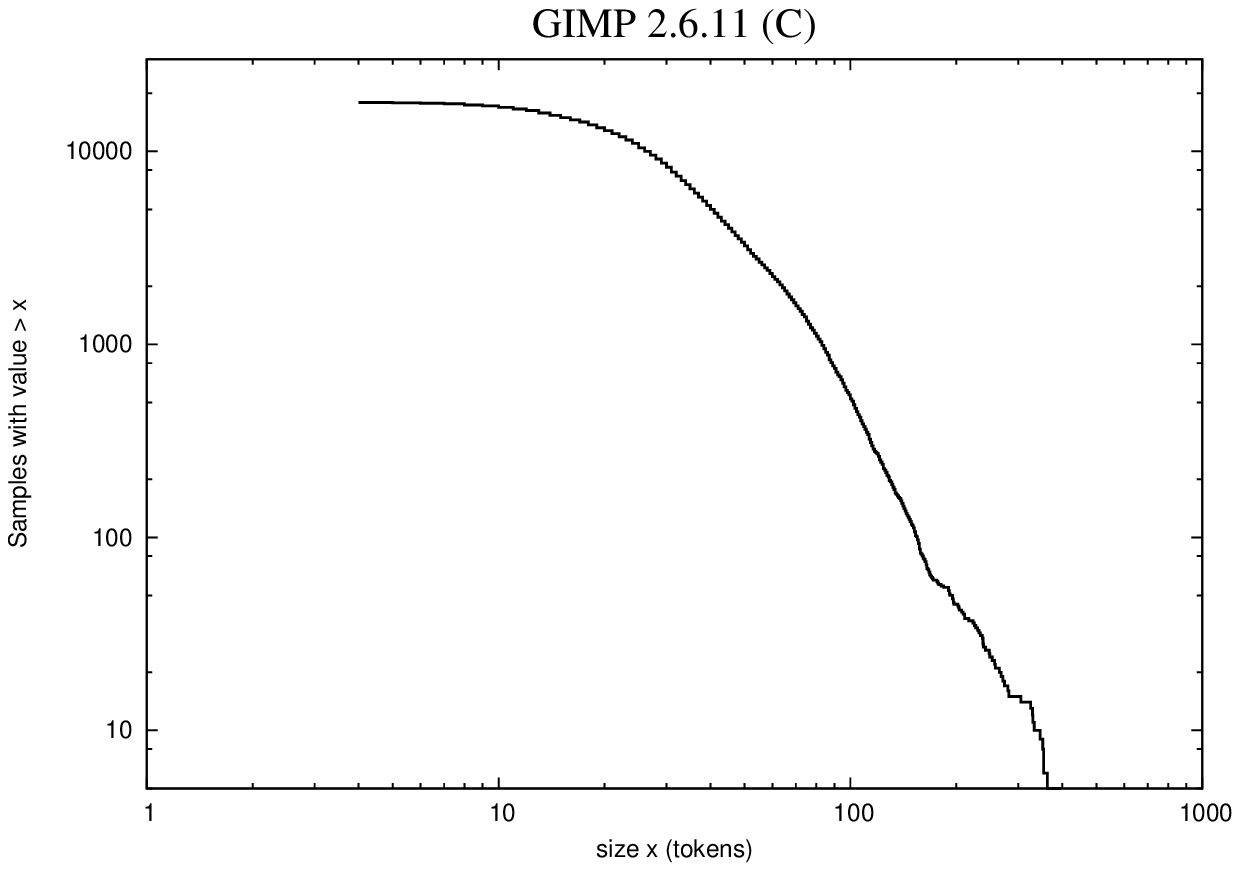,width=0.5\linewidth,clip=} &
\epsfig{file=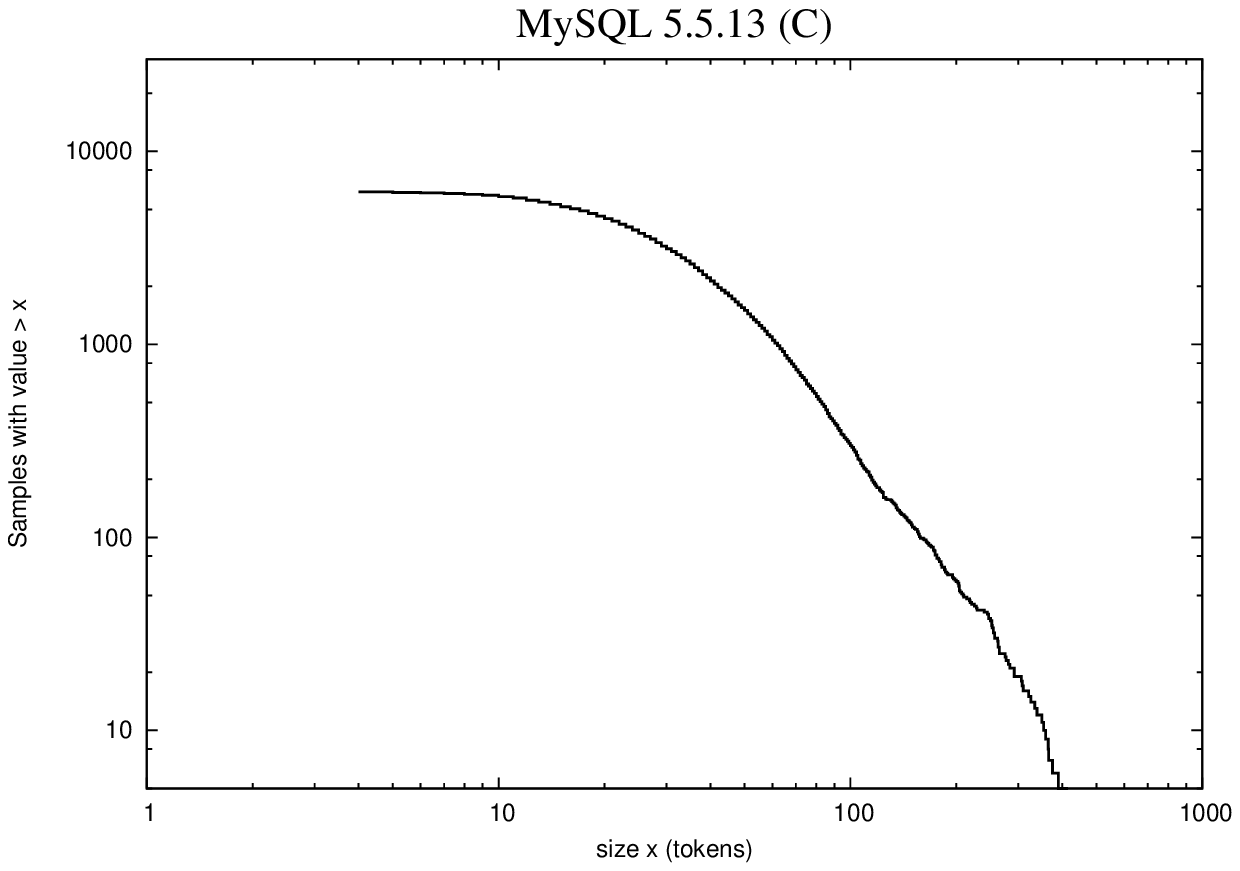,width=0.5\linewidth,clip=} \\
\epsfig{file=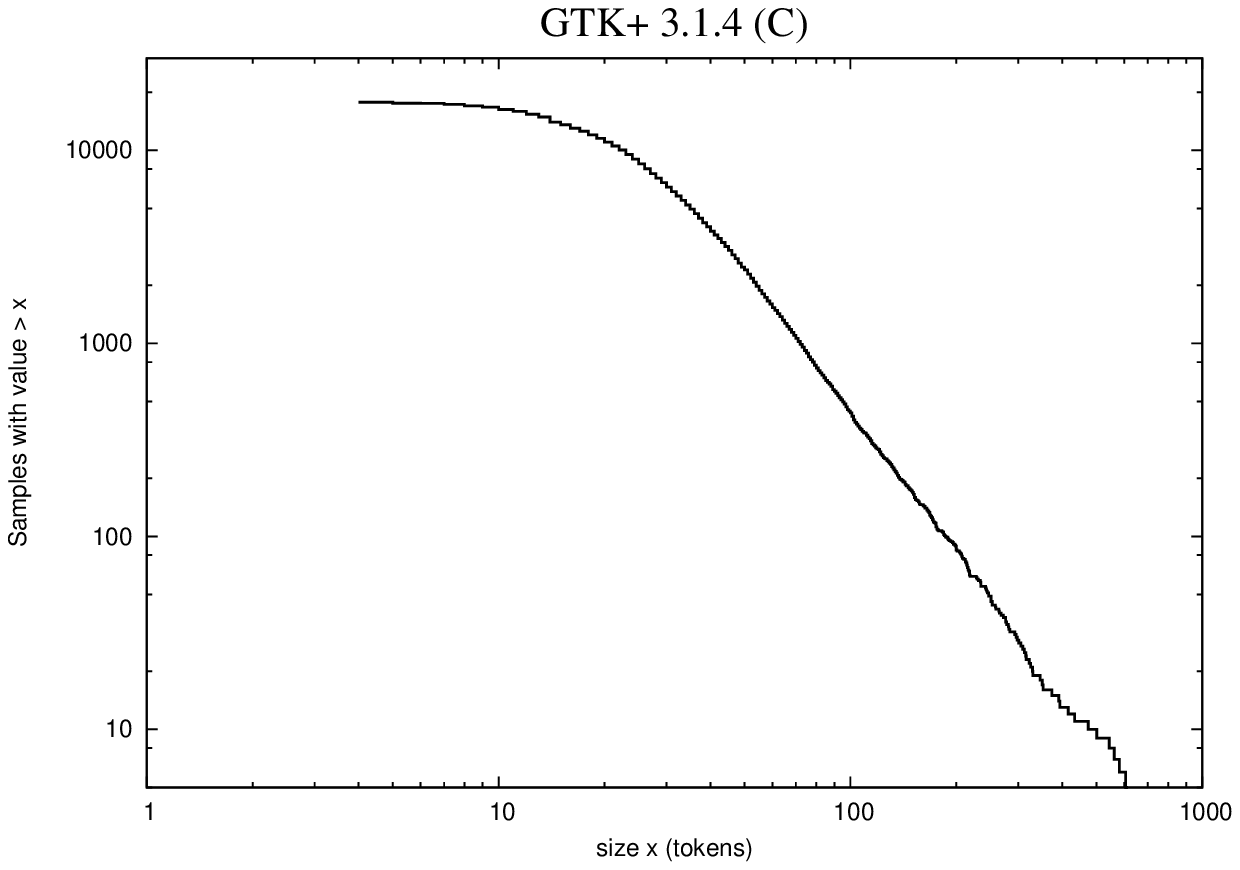,width=0.5\linewidth,clip=} &
\epsfig{file=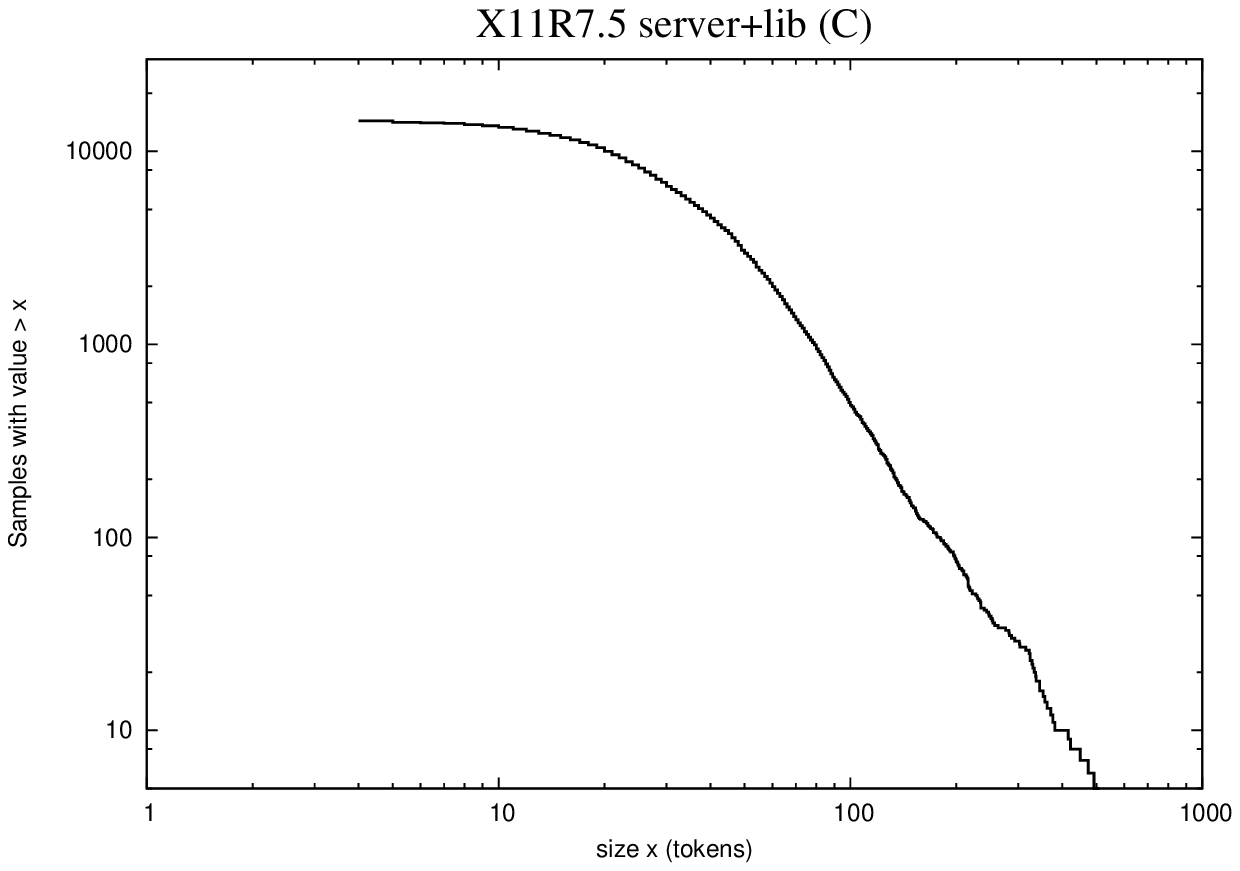,width=0.5\linewidth,clip=} \\
\epsfig{file=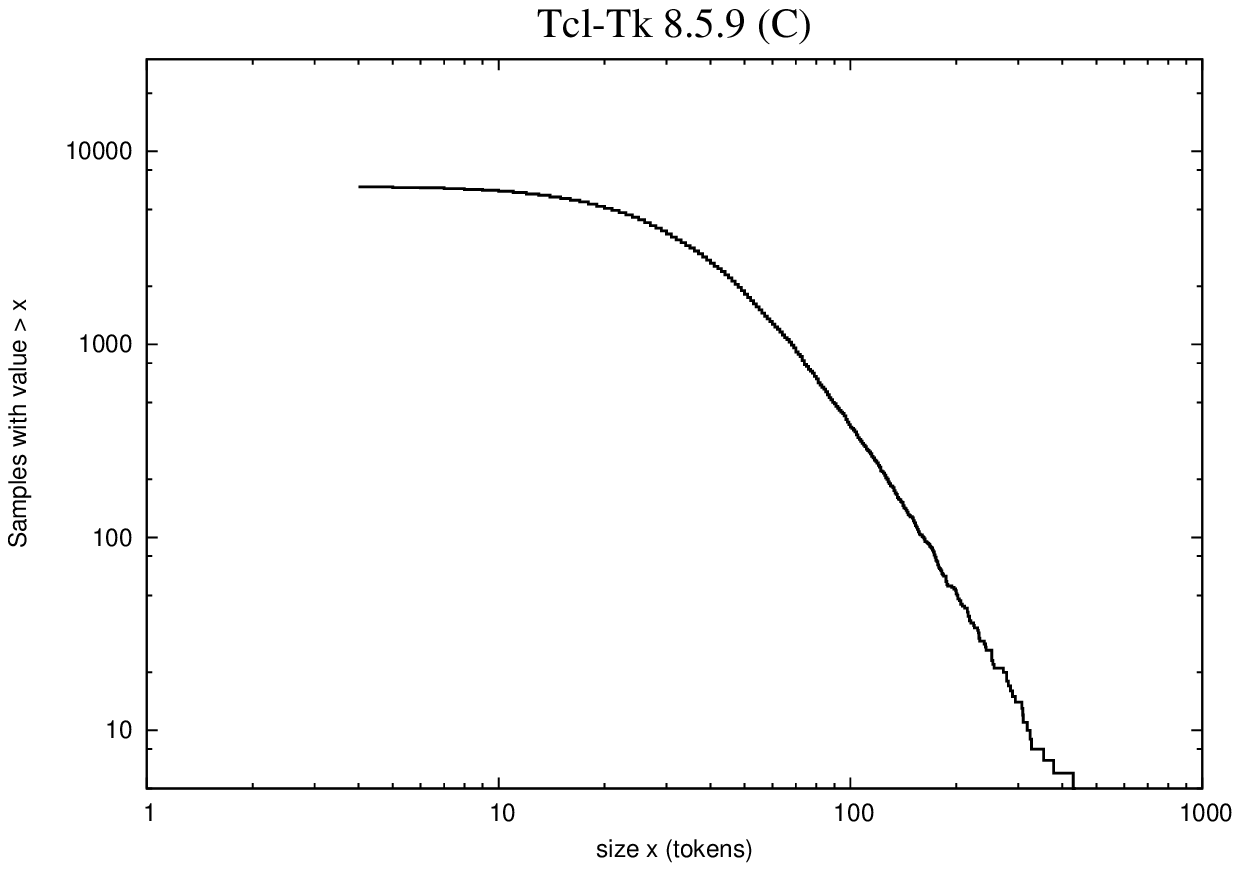,width=0.5\linewidth,clip=} &
\epsfig{file=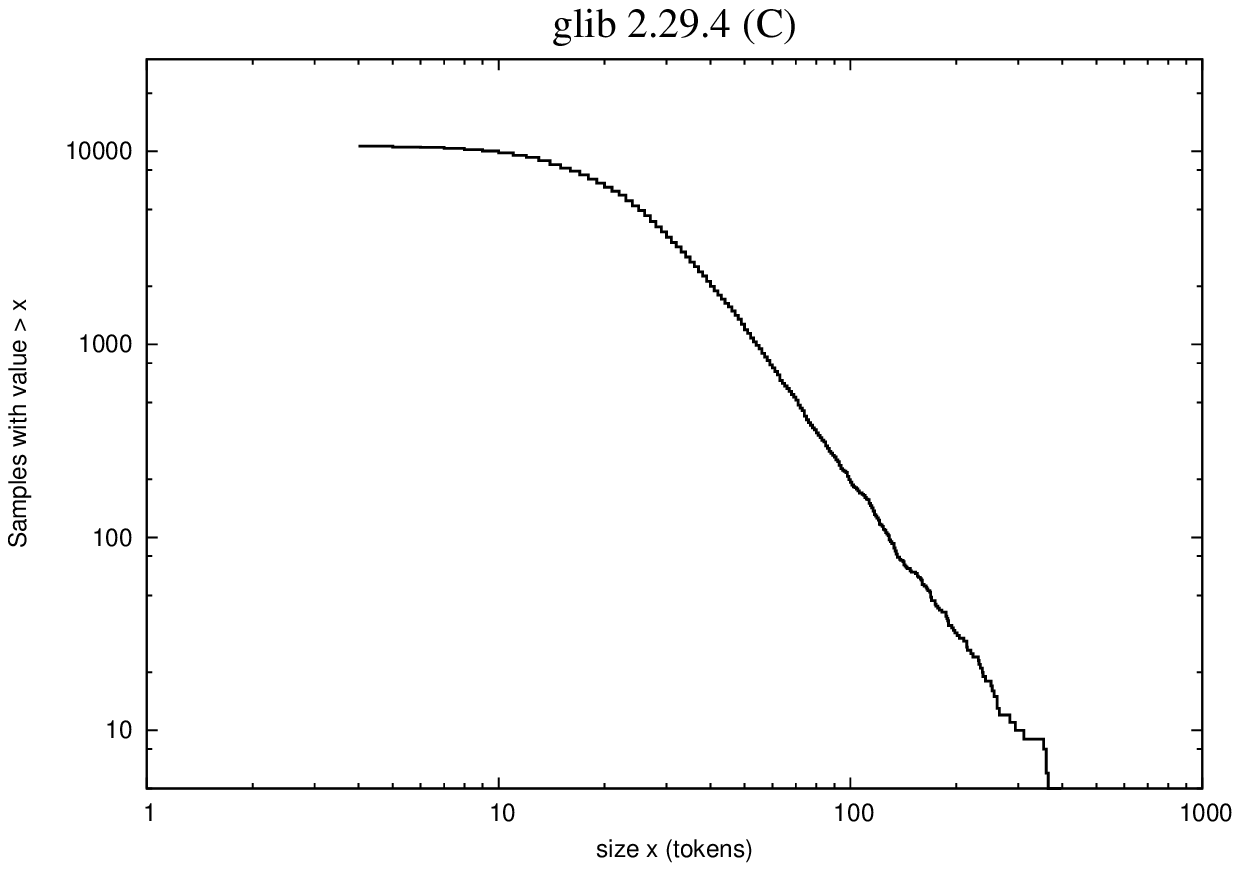,width=0.5\linewidth,clip=} \\
\end{tabular}
\caption{Large C applications in the range 500,000 - 1,100,000 SLOC.  Shown are the latest versions of each of PHP, Mplayer, GIMP, MySQL, GTK, X11R7, TclTk and glib at the time of writing.}
\label{fig:largec}
\end{figure}

It was mentioned earlier that classical statistical mechanics results often remain robust at smaller values of T.  Figure \ref{fig:largec} shows a collage of some of the individual C systems in the range 500,000 - 1,100,000 lines of code all of which show good similarity with the generic model.

\begin{figure}
\centering
\begin{tabular}{cc}
\epsfig{file=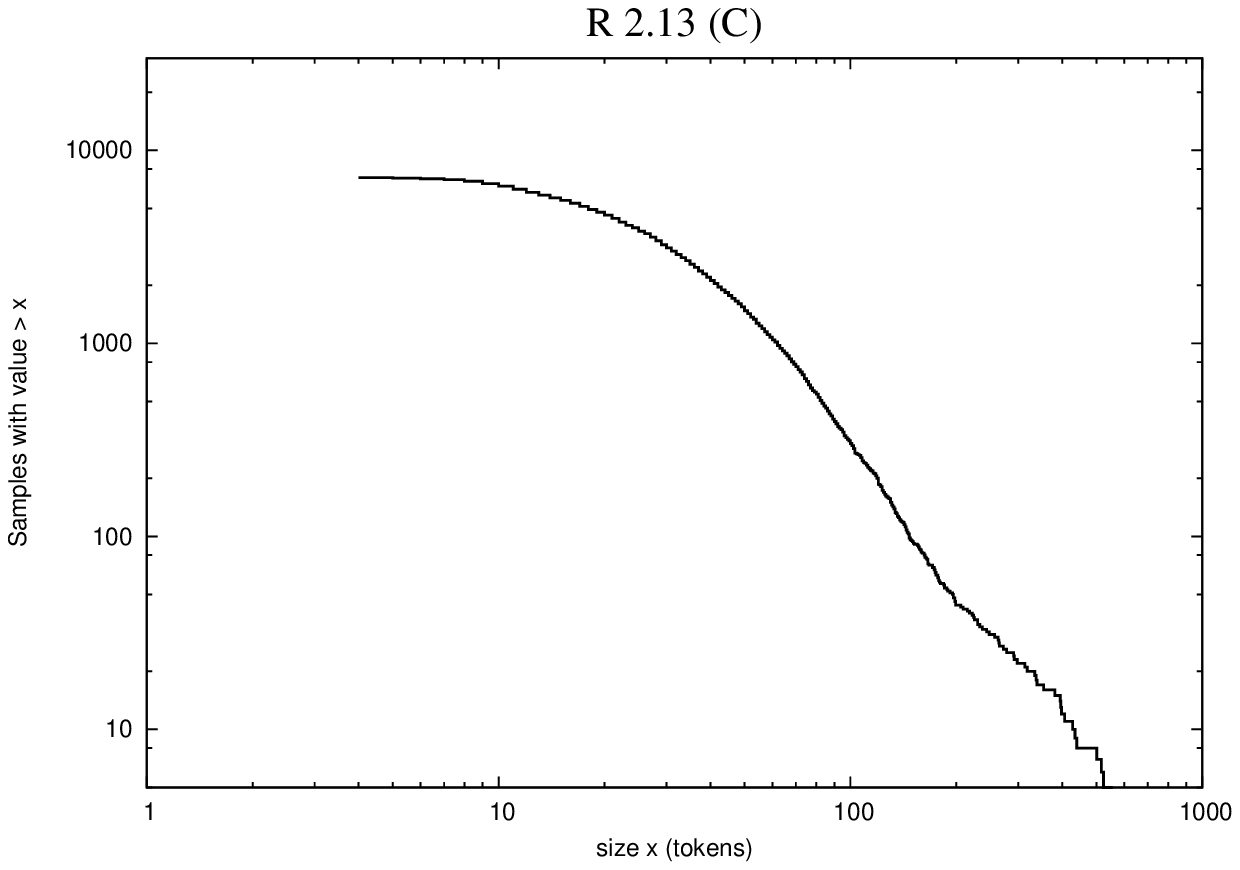,width=0.5\linewidth,clip=} &
\epsfig{file=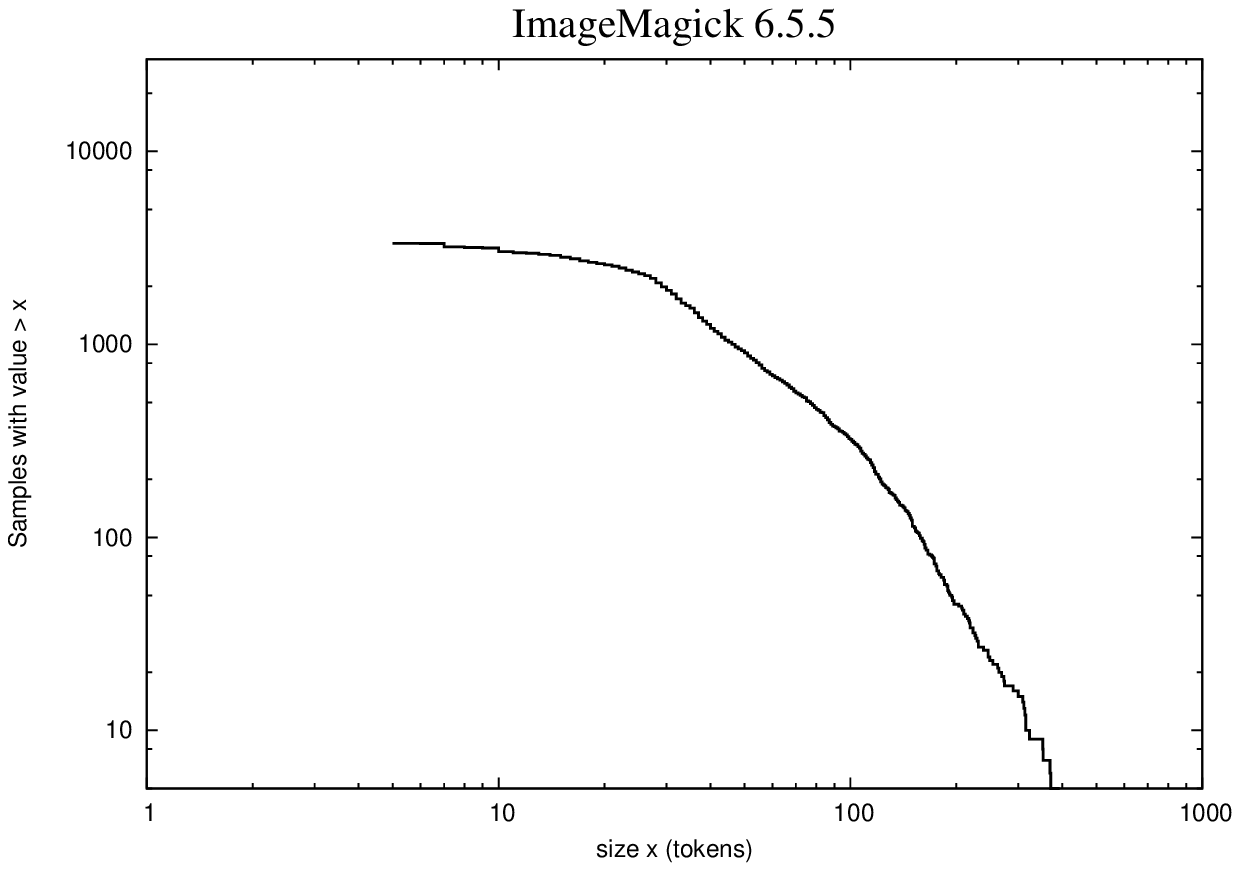,width=0.5\linewidth,clip=} \\
\epsfig{file=gimp.eps,width=0.5\linewidth,clip=} &
\epsfig{file=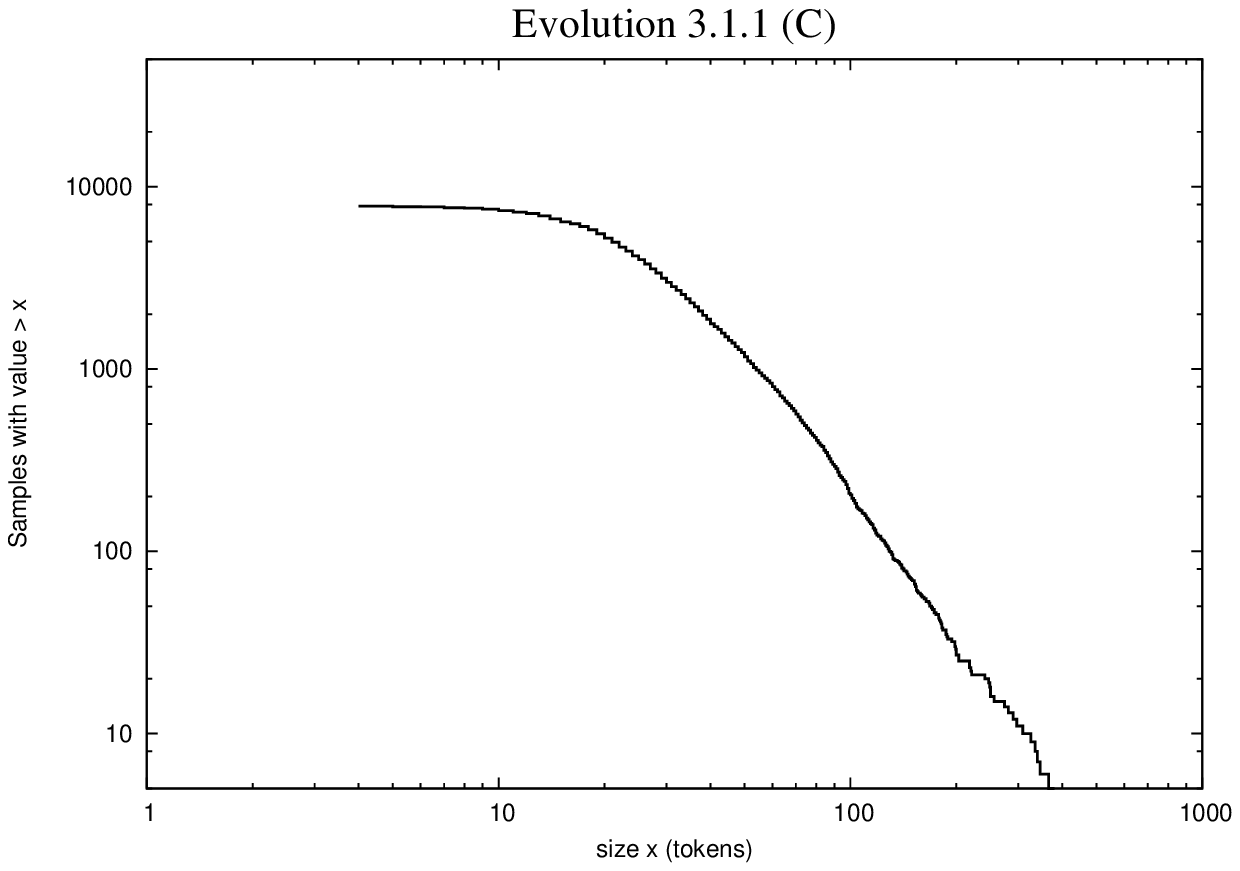,width=0.5\linewidth,clip=} \\
\epsfig{file=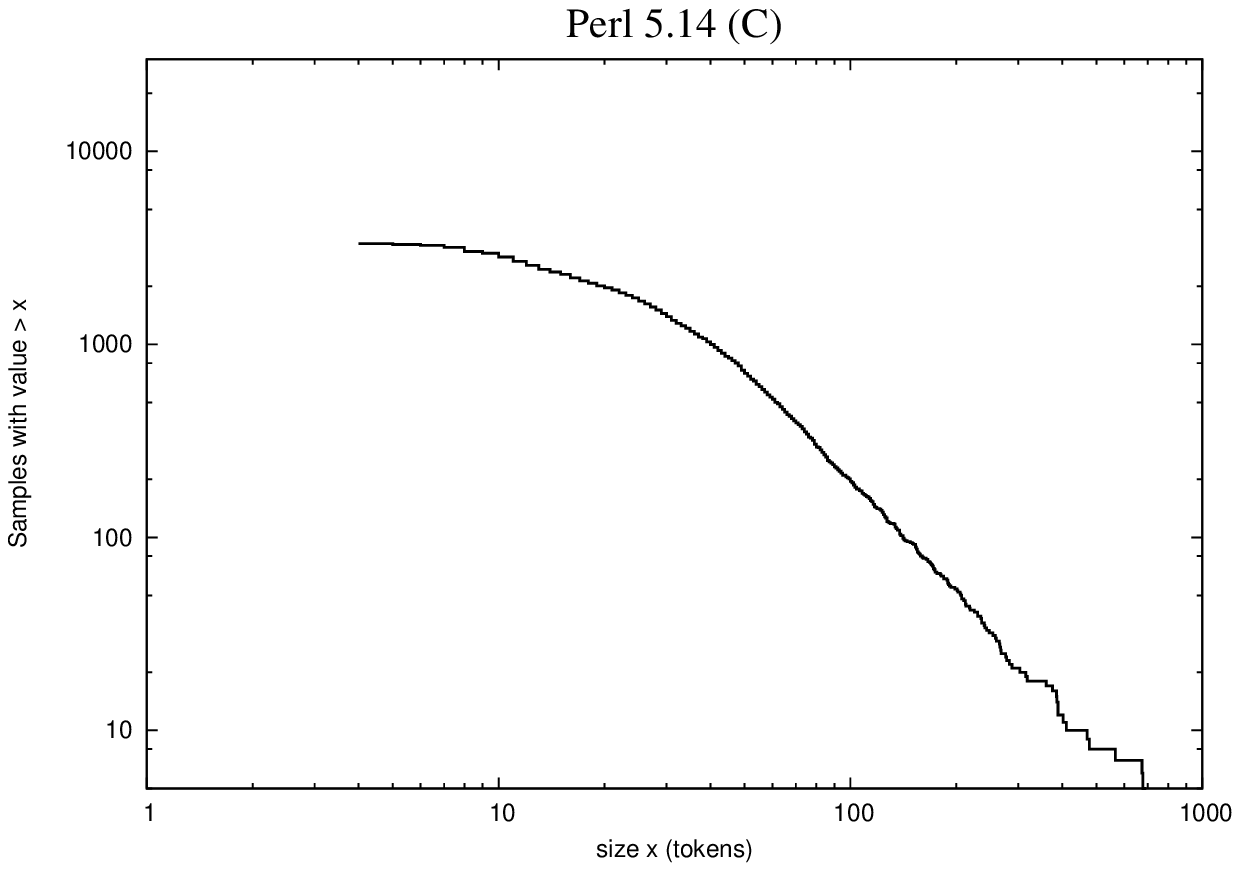,width=0.5\linewidth,clip=} &
\epsfig{file=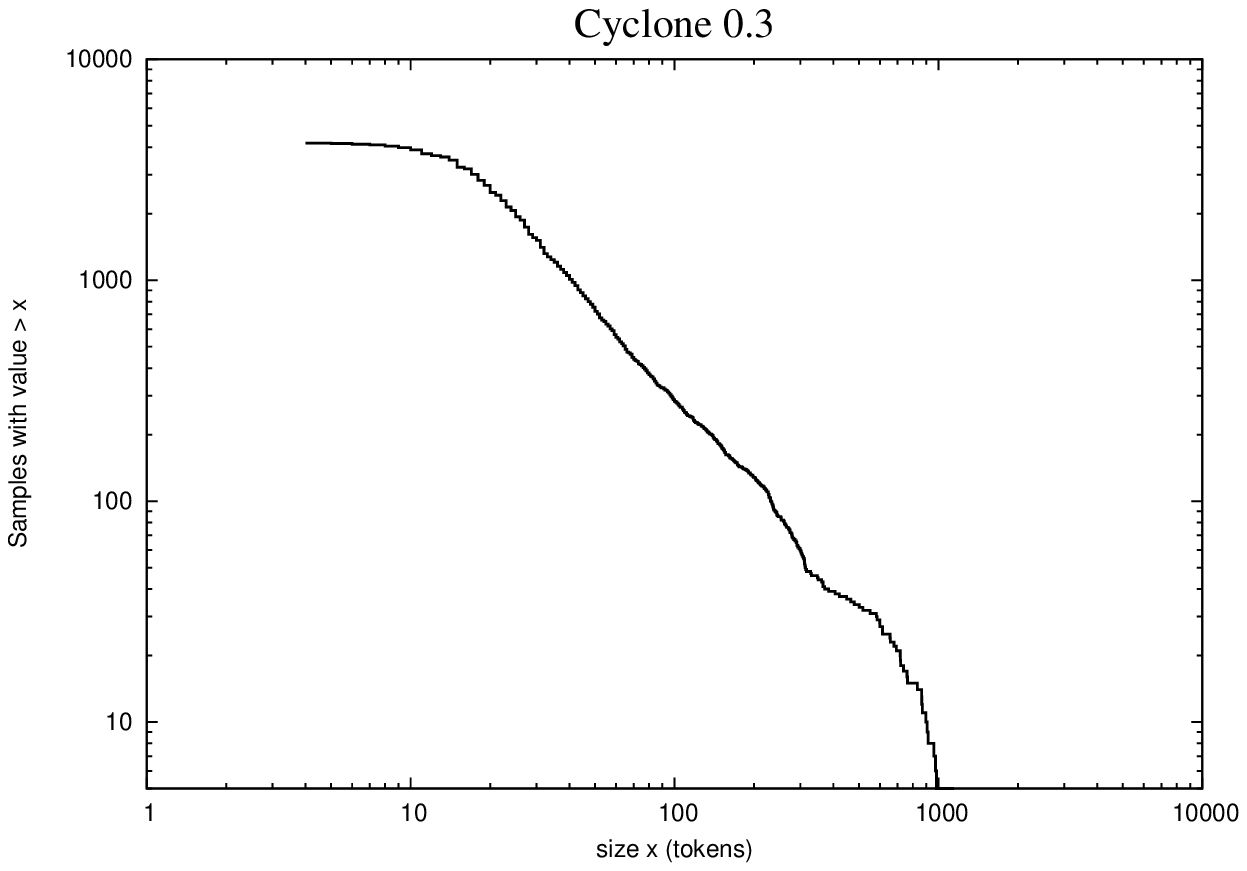,width=0.5\linewidth,clip=} \\
\epsfig{file=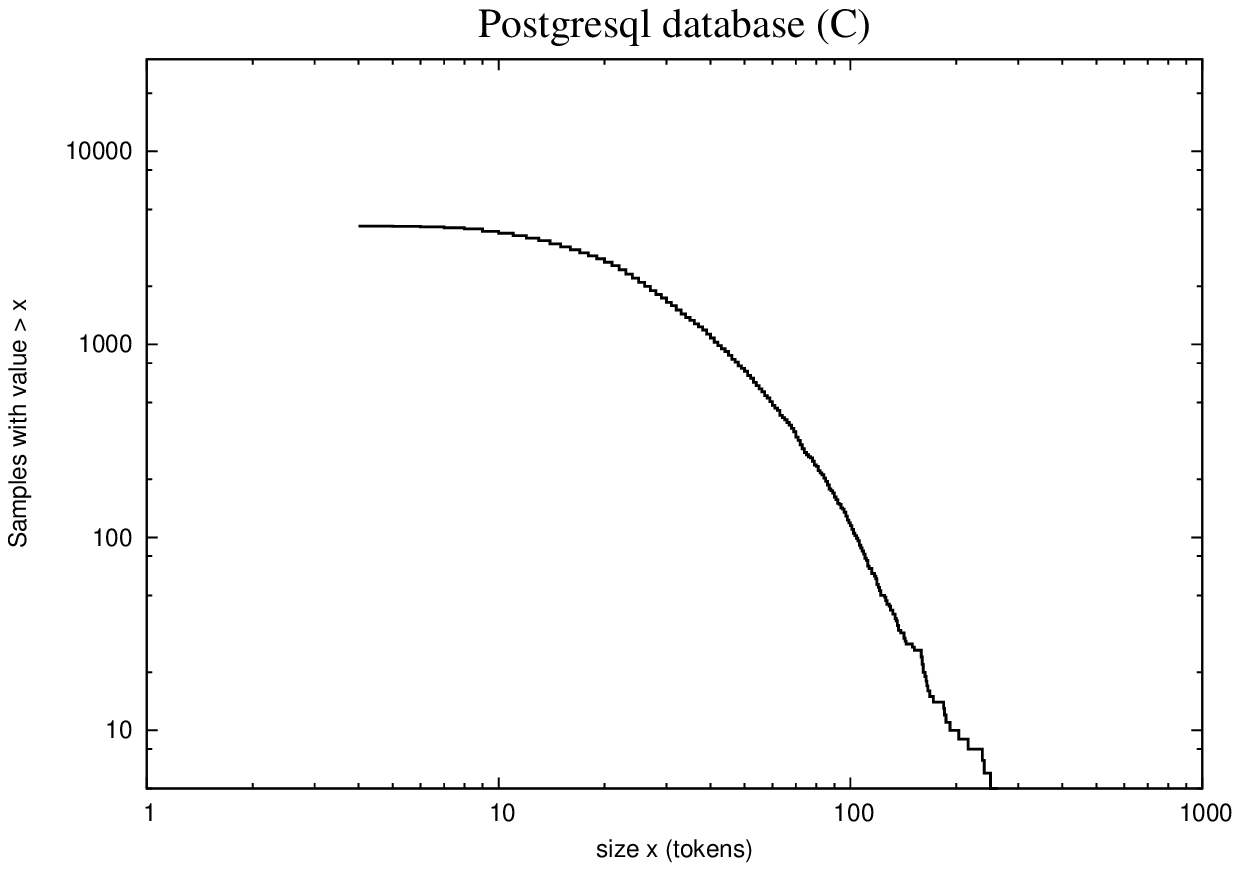,width=0.5\linewidth,clip=} &
\epsfig{file=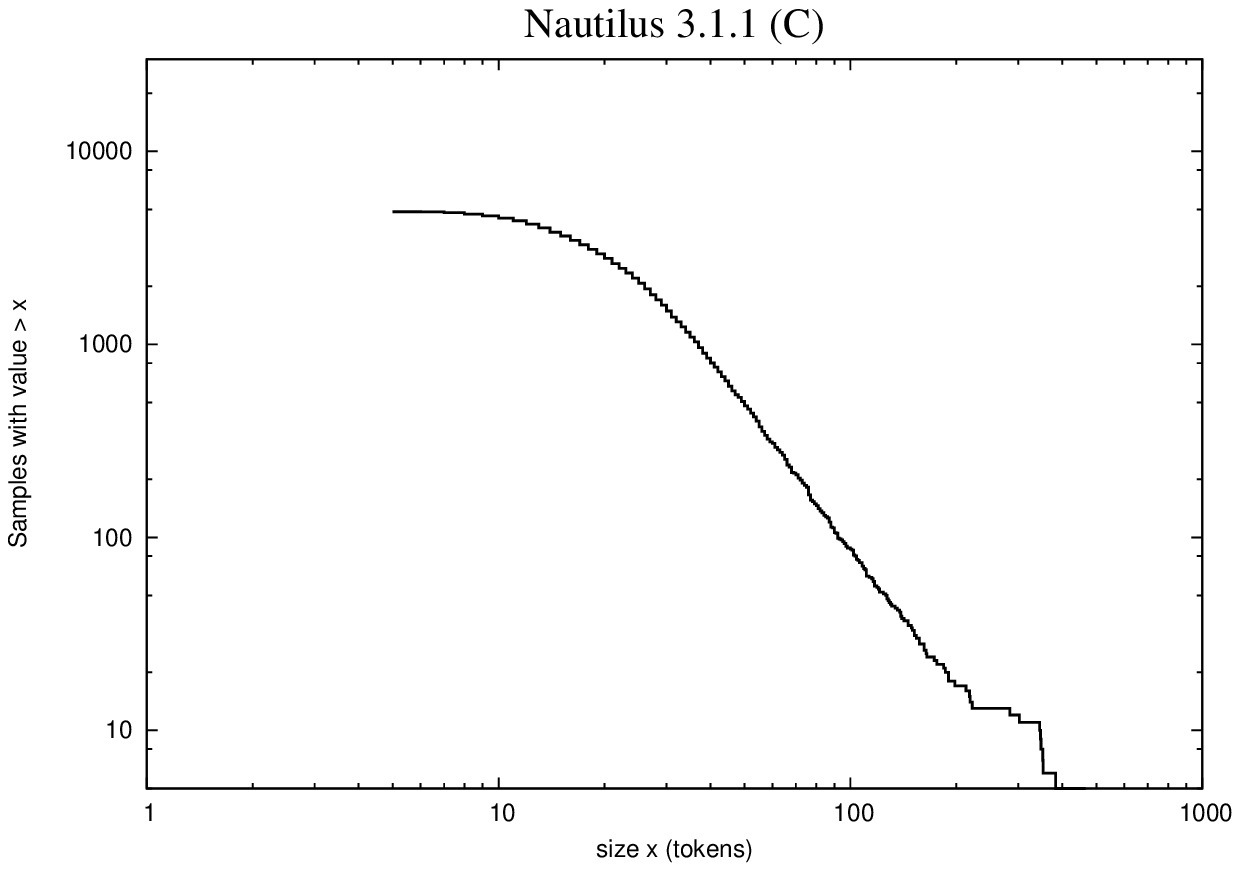,width=0.5\linewidth,clip=} \\
\end{tabular}
\caption{Medium C applications in the range 150,000 - 400,000 SLOC.}
\label{fig:mediumc}
\end{figure}

\begin{figure}
\centering
\begin{tabular}{cc}
\epsfig{file=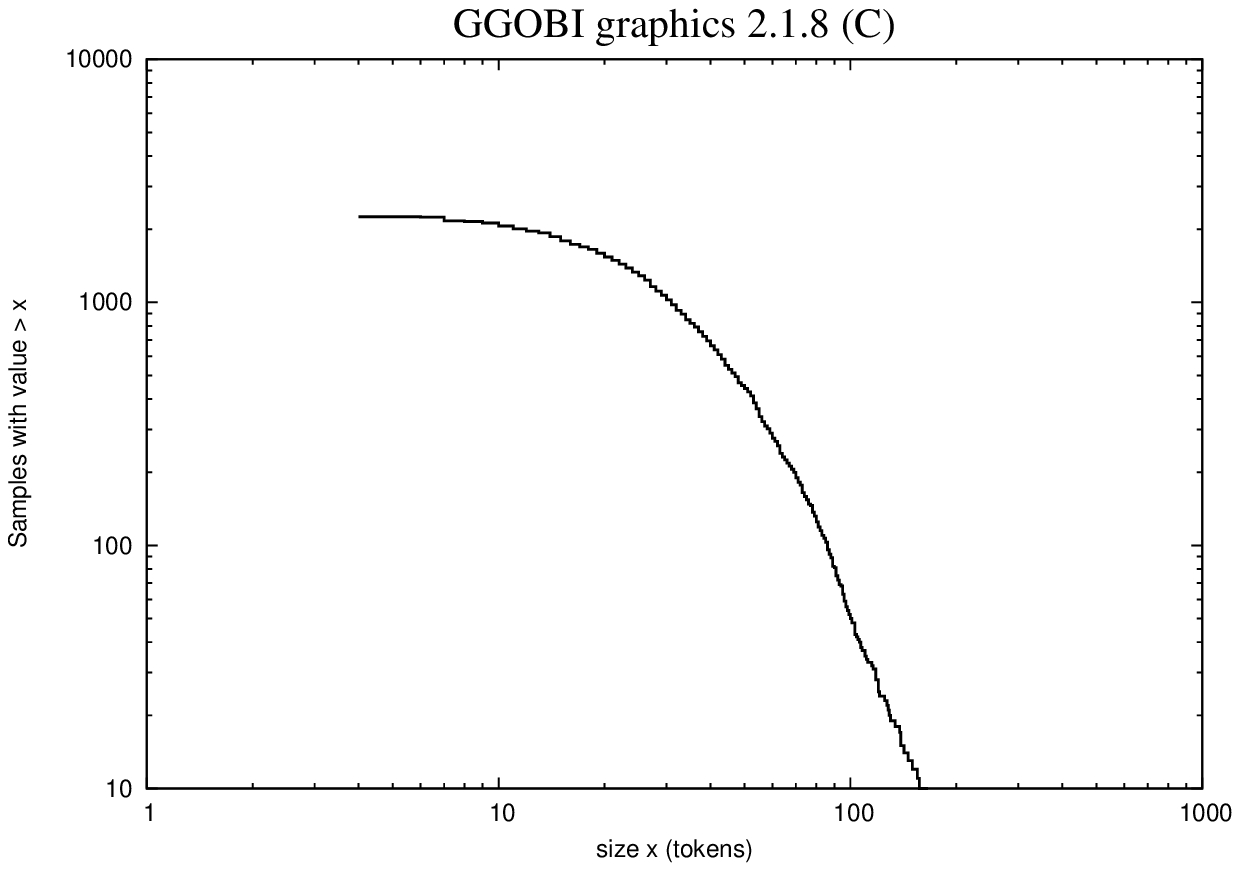,width=0.5\linewidth,clip=} &
\epsfig{file=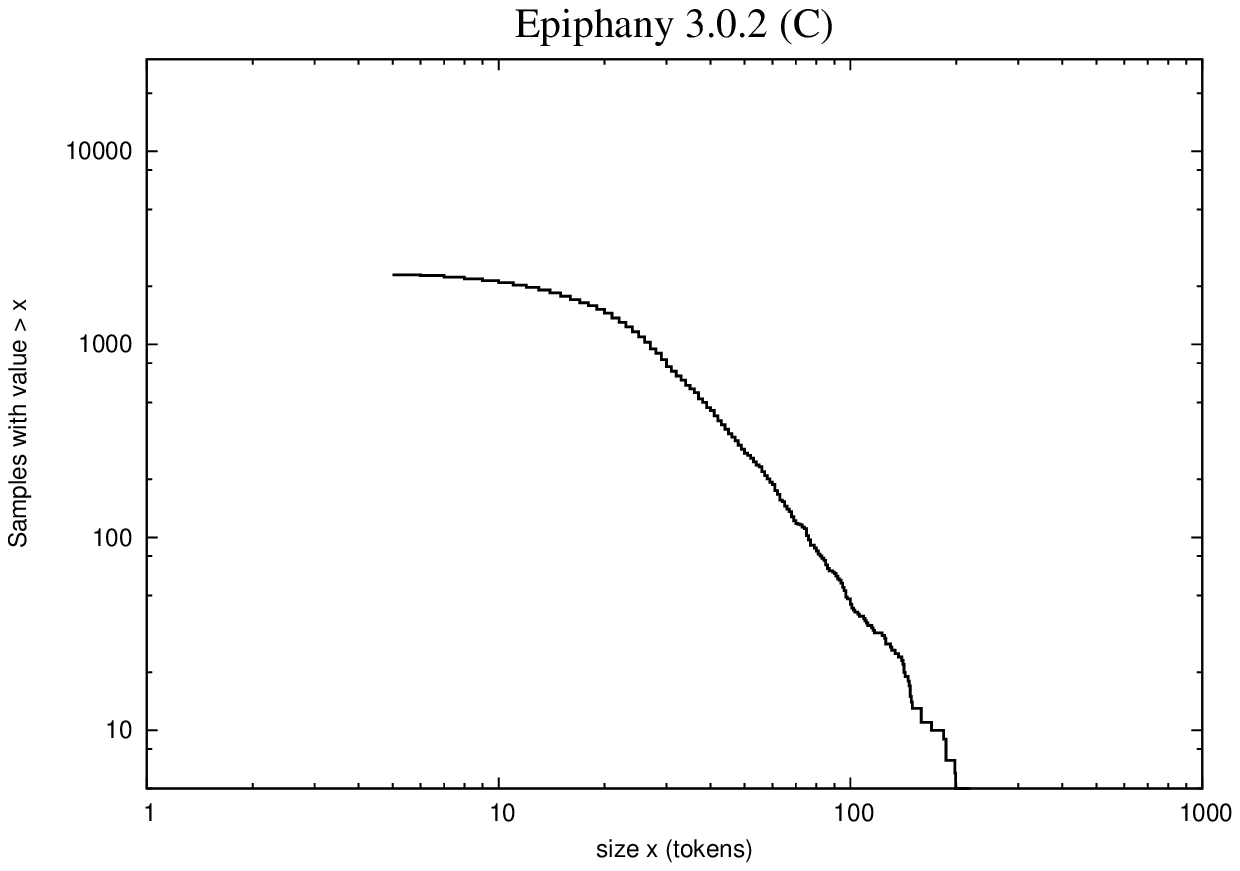,width=0.5\linewidth,clip=} \\
\epsfig{file=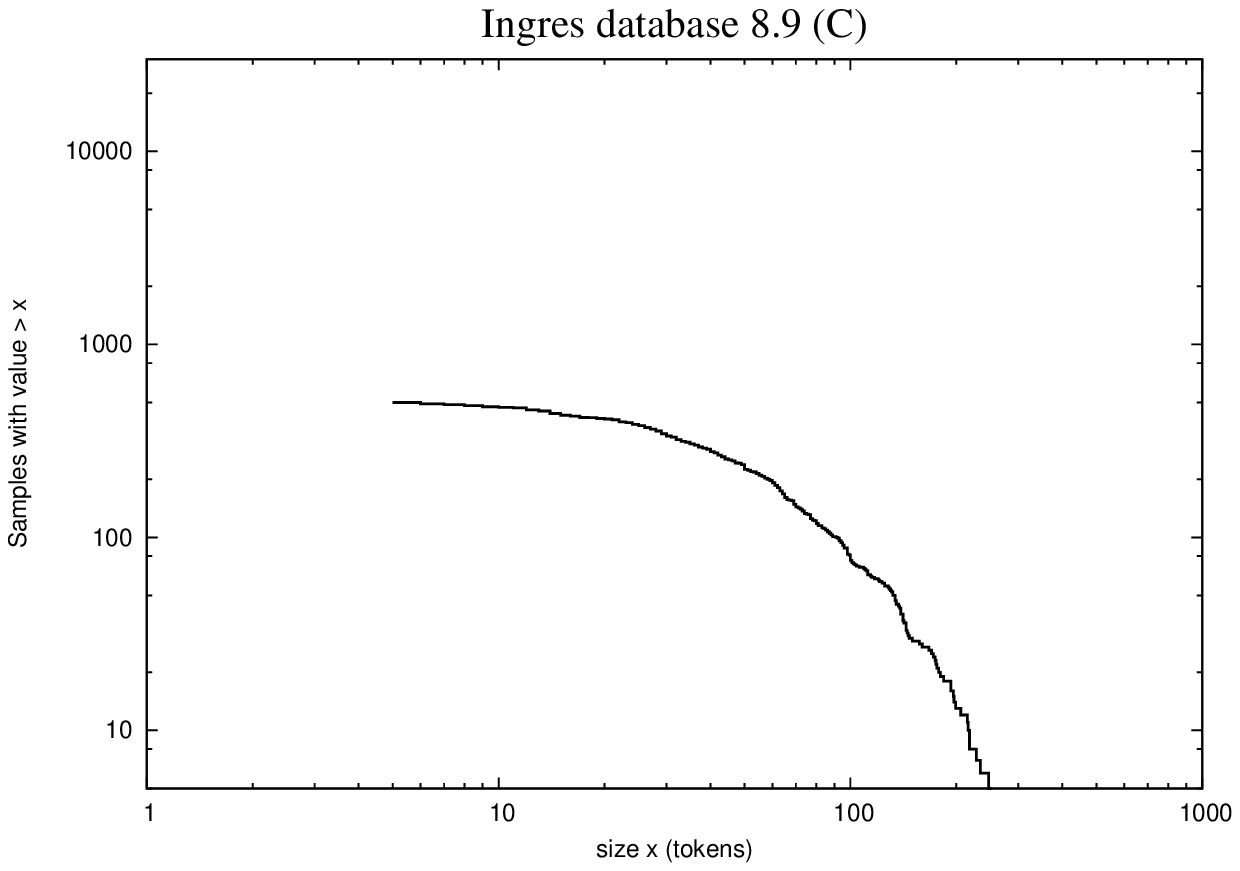,width=0.5\linewidth,clip=} &
\epsfig{file=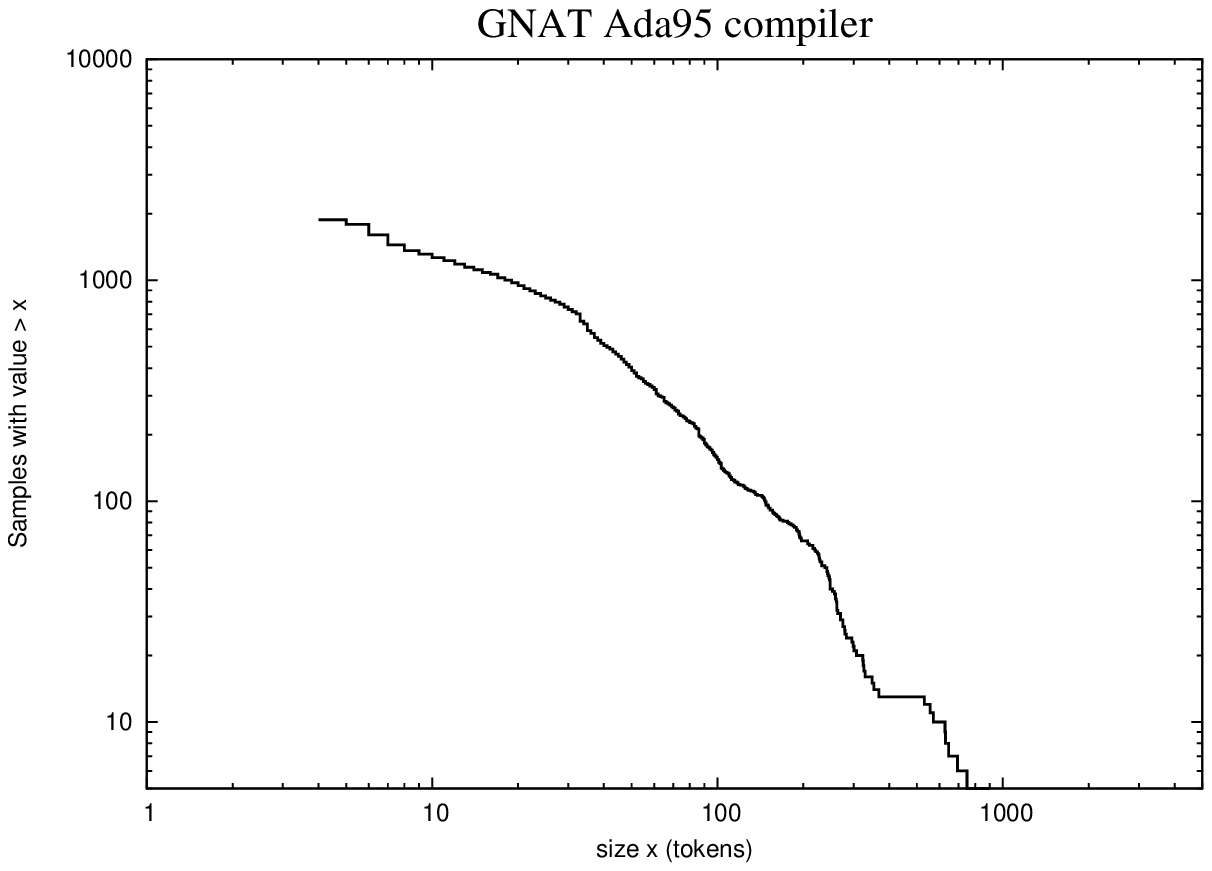,width=0.5\linewidth,clip=} \\
\epsfig{file=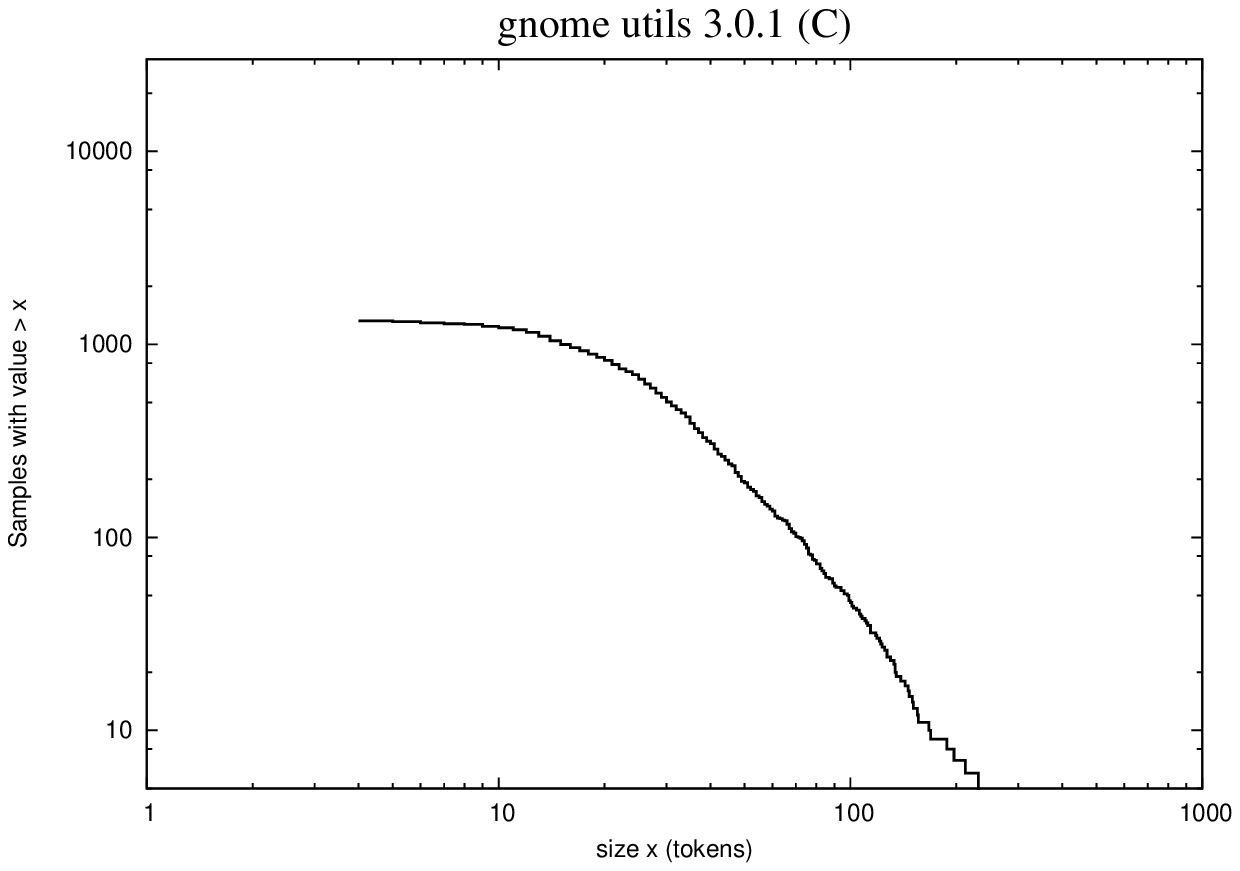,width=0.5\linewidth,clip=} &
\epsfig{file=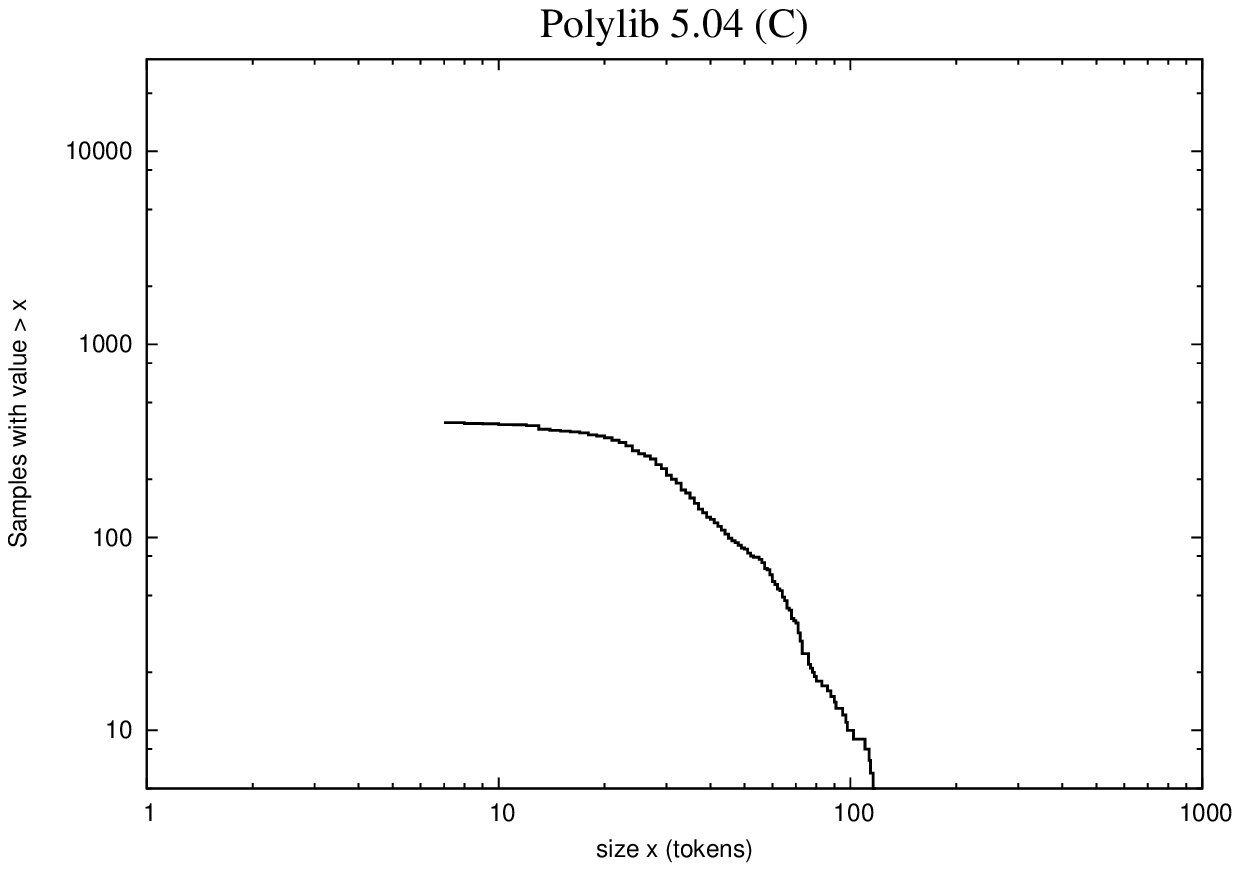,width=0.5\linewidth,clip=} \\
\epsfig{file=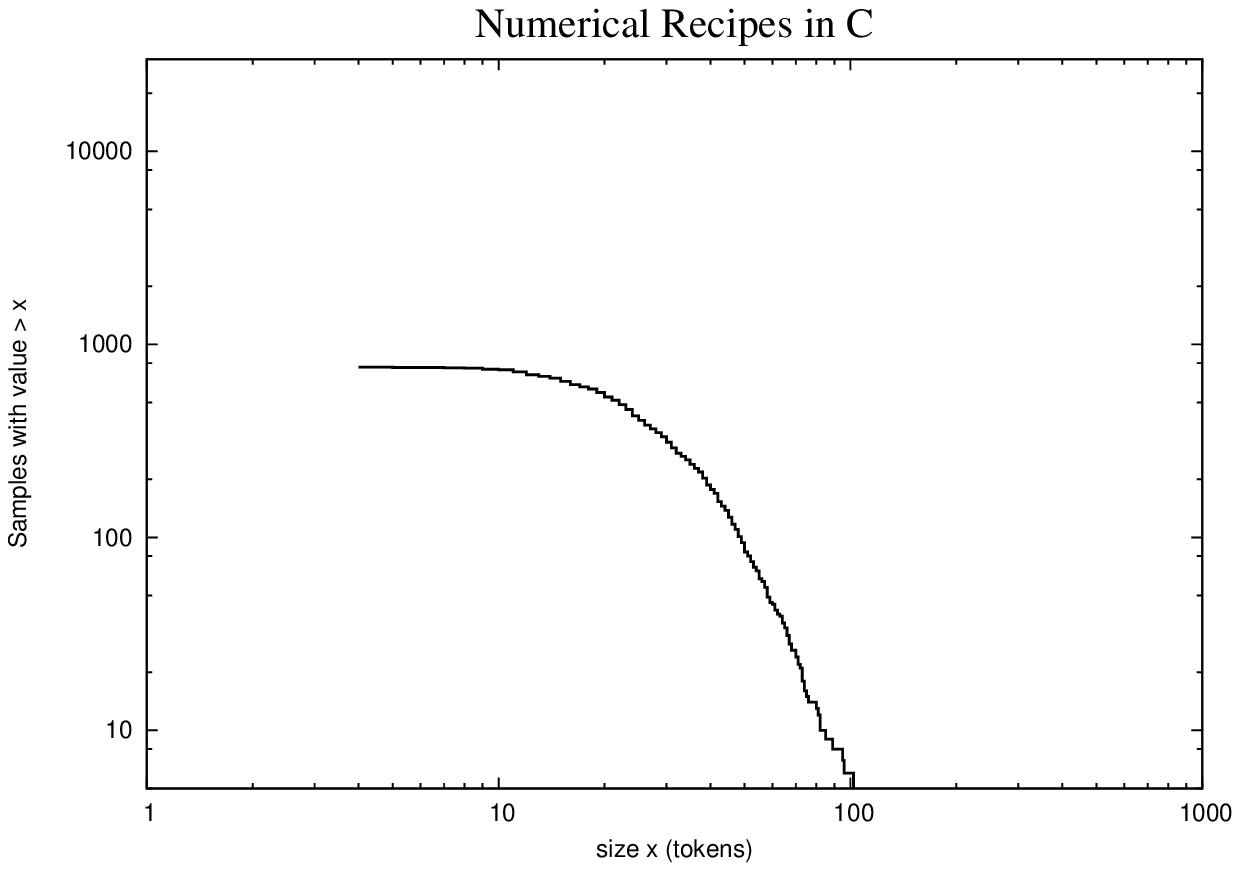,width=0.5\linewidth,clip=} &
\epsfig{file=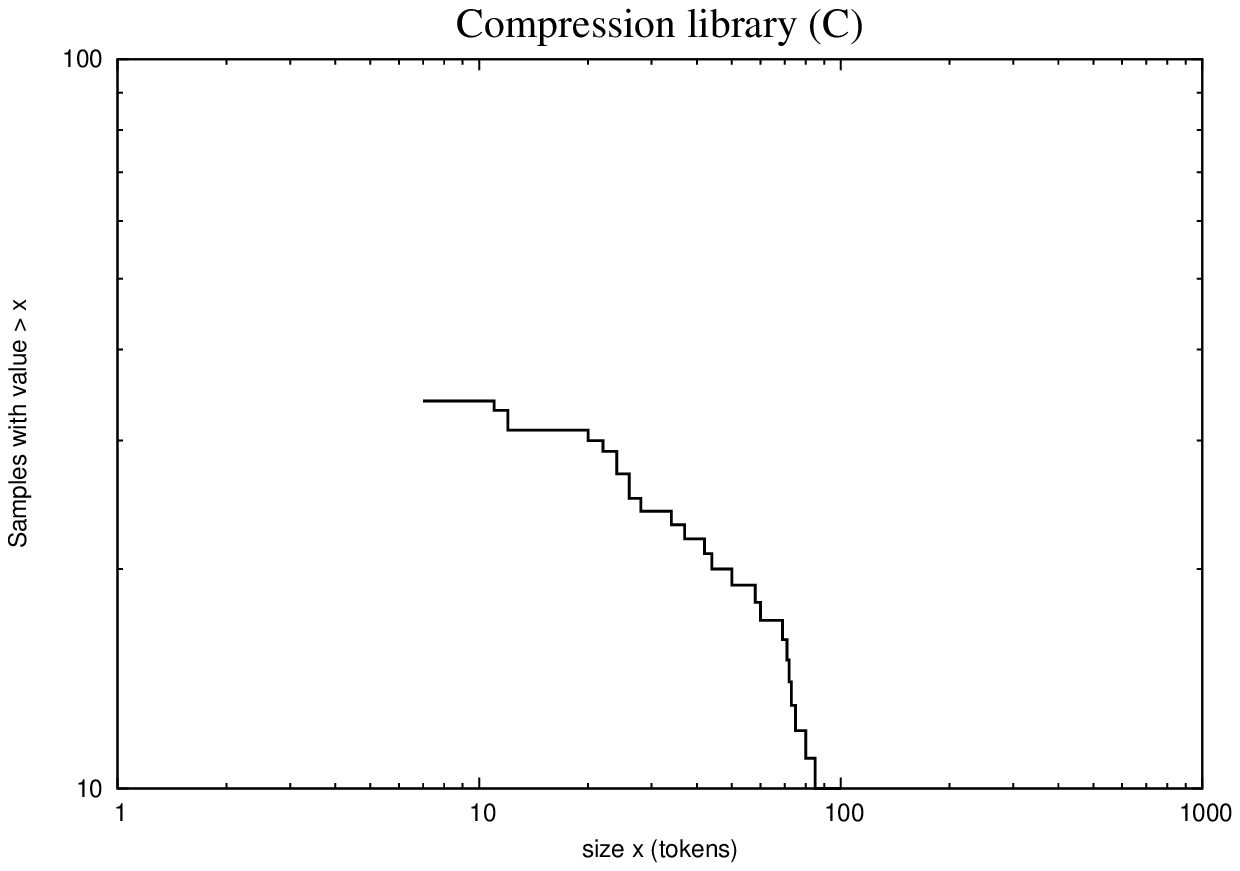,width=0.5\linewidth,clip=} \\
\end{tabular}
\caption{Small C applications in the range 8,000 - 90,000 SLOC.}
\label{fig:smallc}
\end{figure}

As can be seen by studying the animation at http://www.leshatton.org/wp-content/uploads/2012/01/animate.gif, the generic shape of (\ref{eq:pwrlaw}) appears fairly early on, certainly within the first 1\% of the total data represented by Figure \ref{fig:universe}.  To give some of idea of medium and small systems, Figures \ref{fig:mediumc} and \ref{fig:smallc} show a collage of individual systems in the ranges 150,000 - 400,000 lines of code and 8,000 - 90,000 lines of code respectively.  

\begin{figure}
\centering
\begin{tabular}{cc}
\epsfig{file=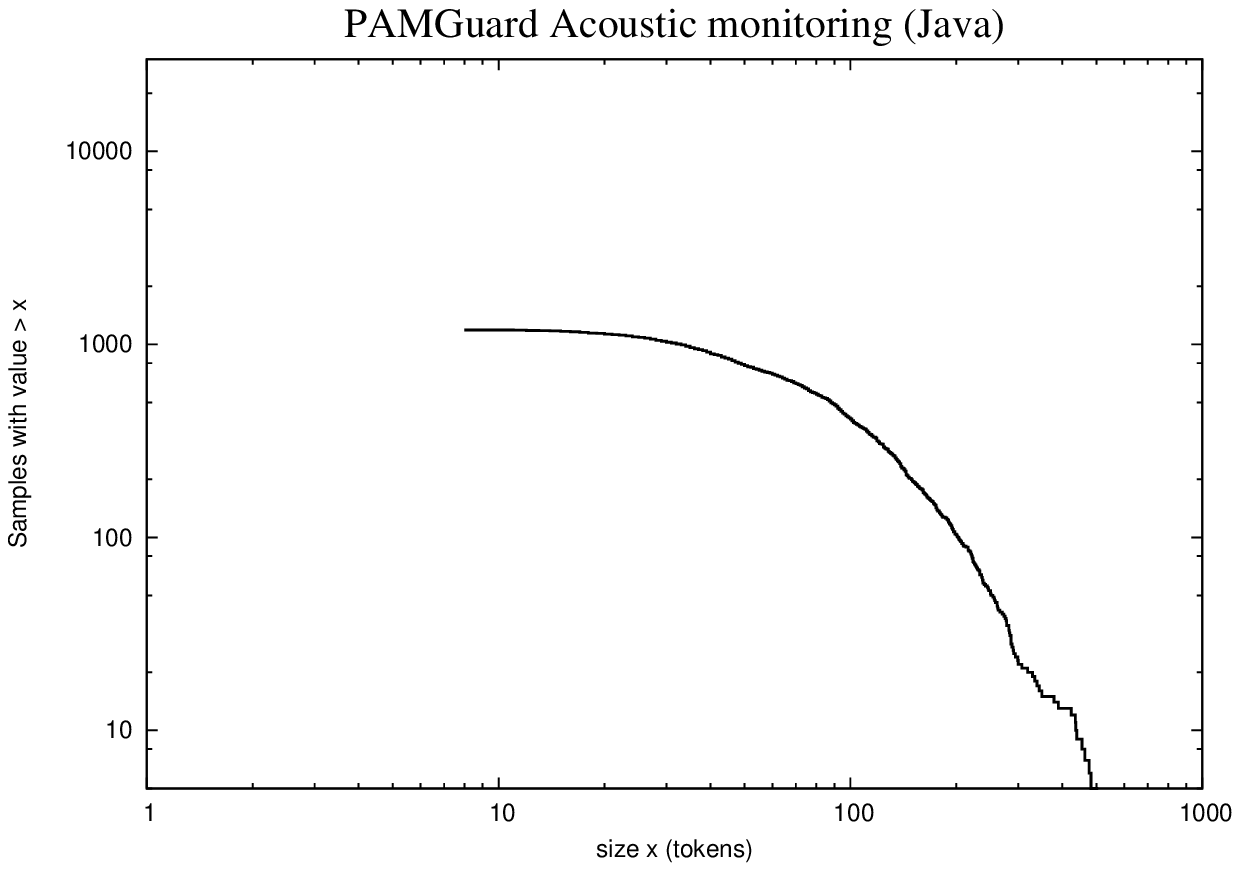,width=0.5\linewidth,clip=} &
\epsfig{file=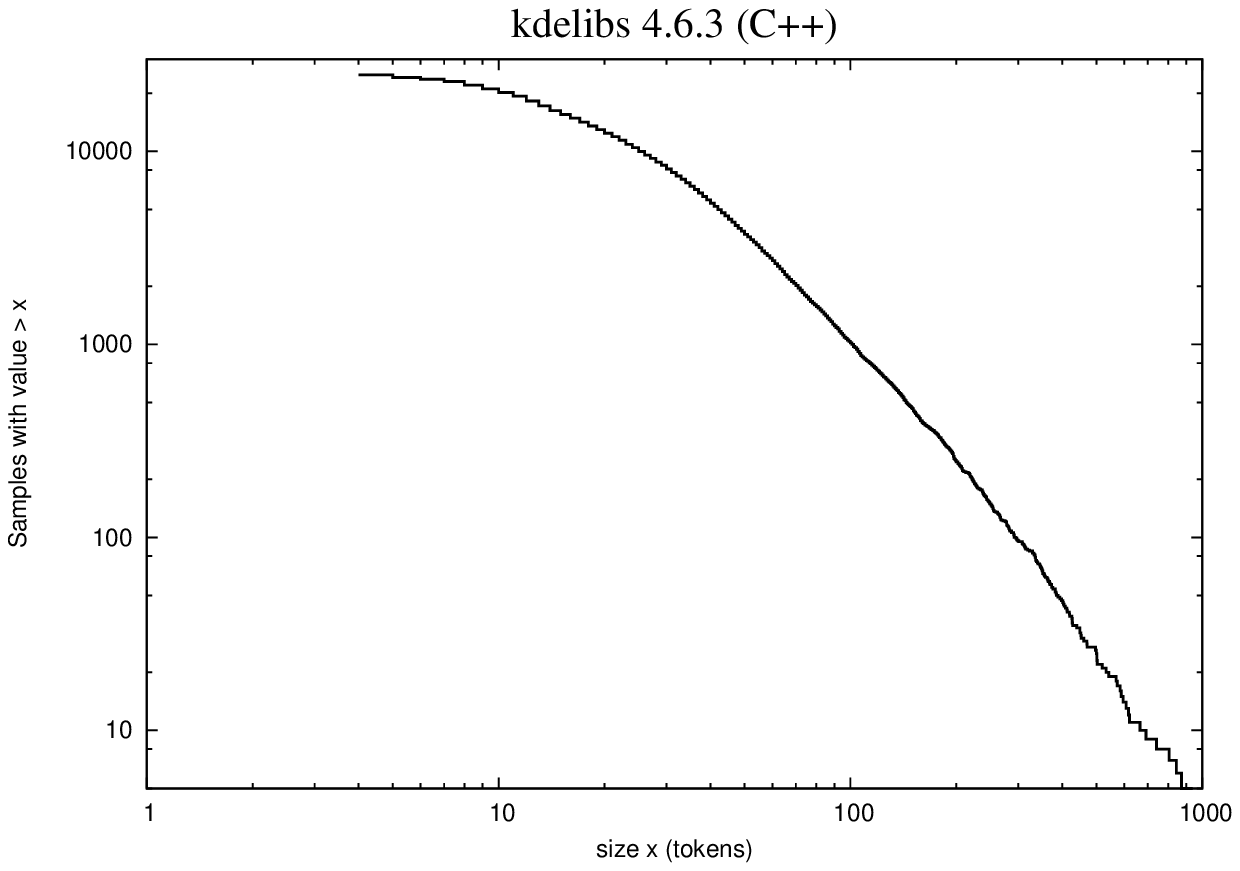,width=0.5\linewidth,clip=} \\
\epsfig{file=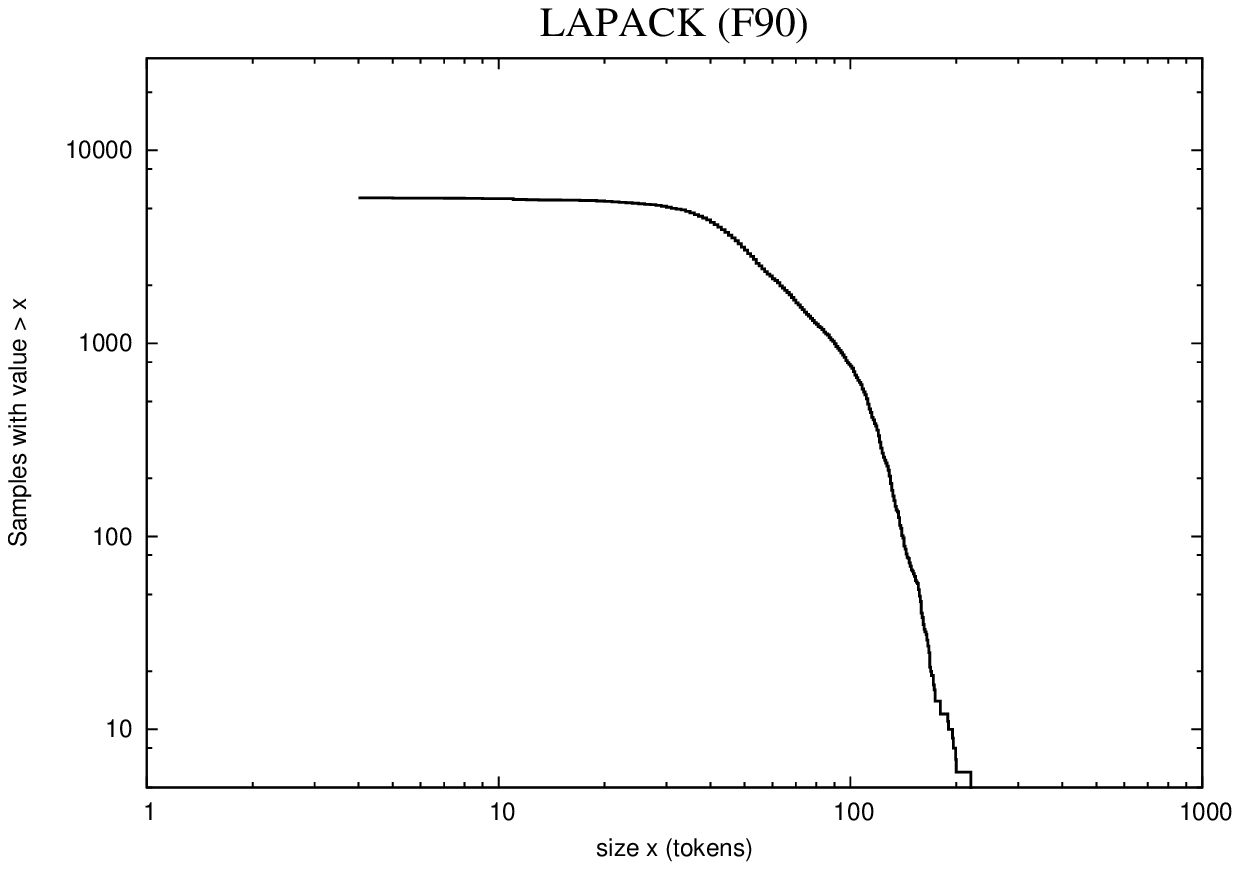,width=0.5\linewidth,clip=} &
\epsfig{file=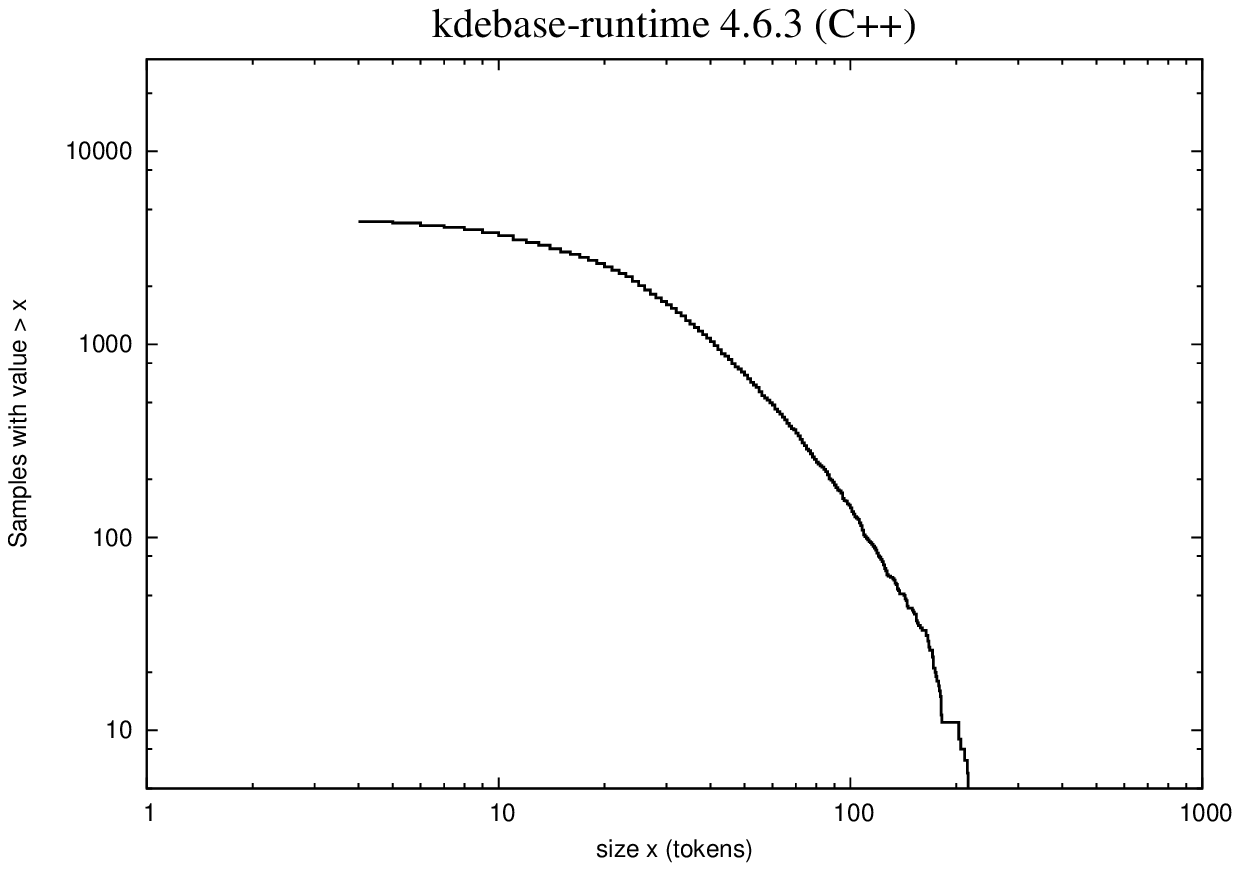,width=0.5\linewidth,clip=} \\
\epsfig{file=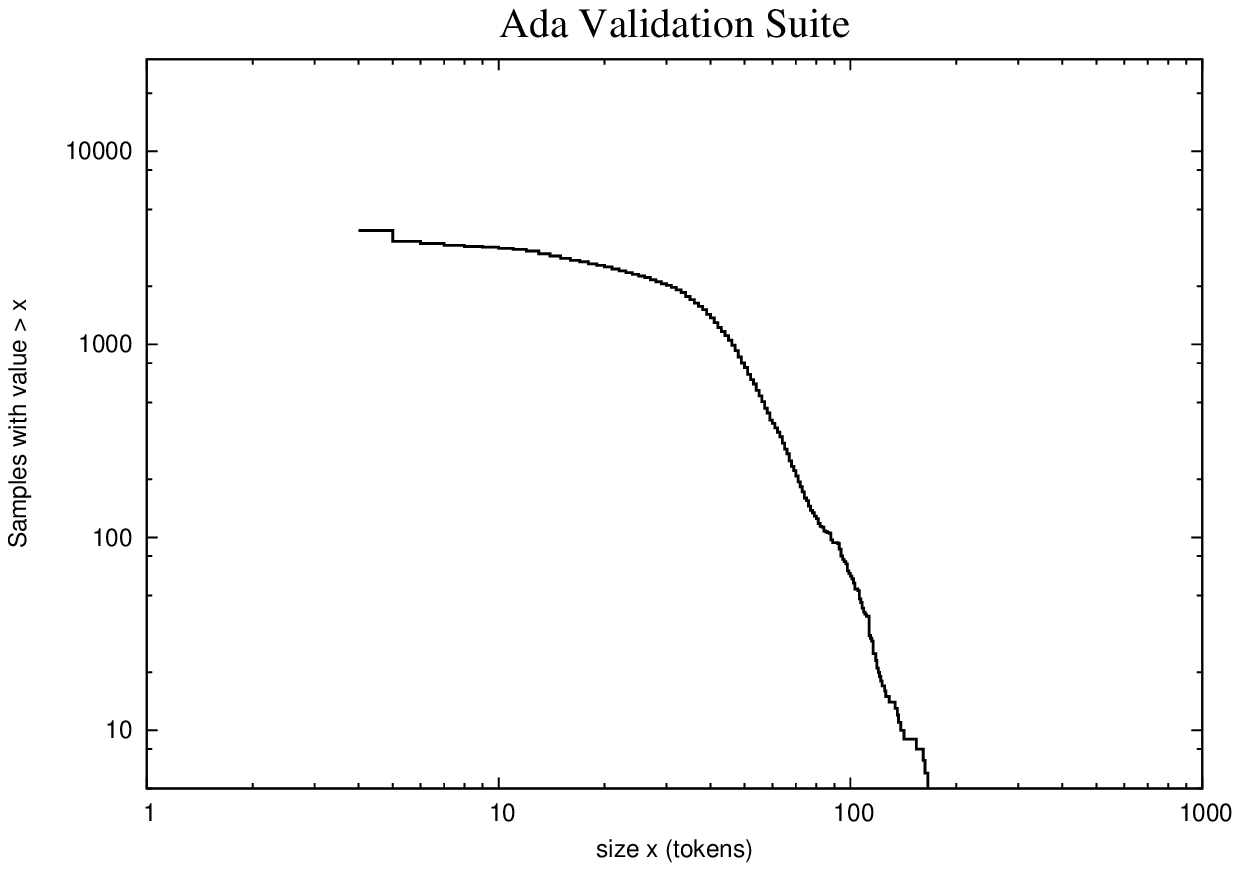,width=0.5\linewidth,clip=} &
\epsfig{file=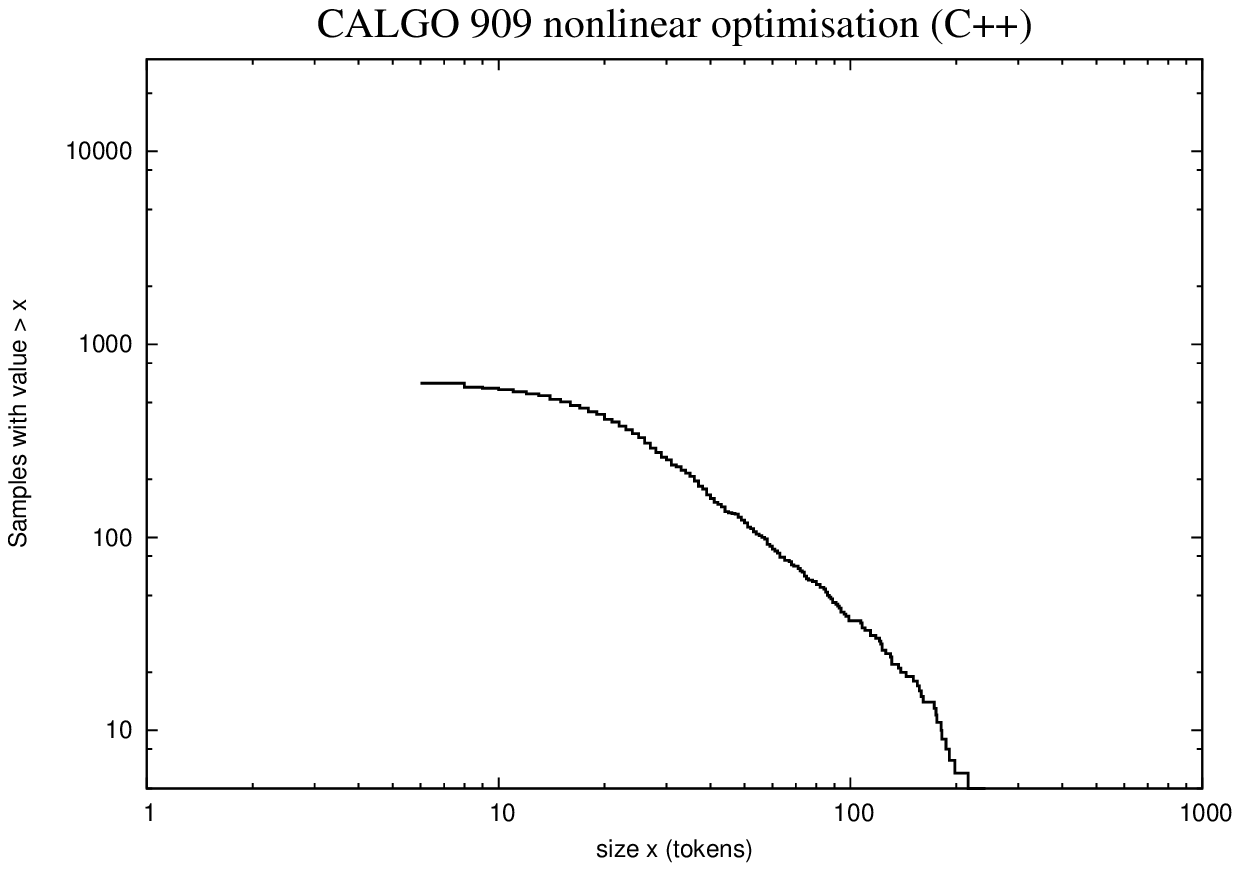,width=0.5\linewidth,clip=} \\
\epsfig{file=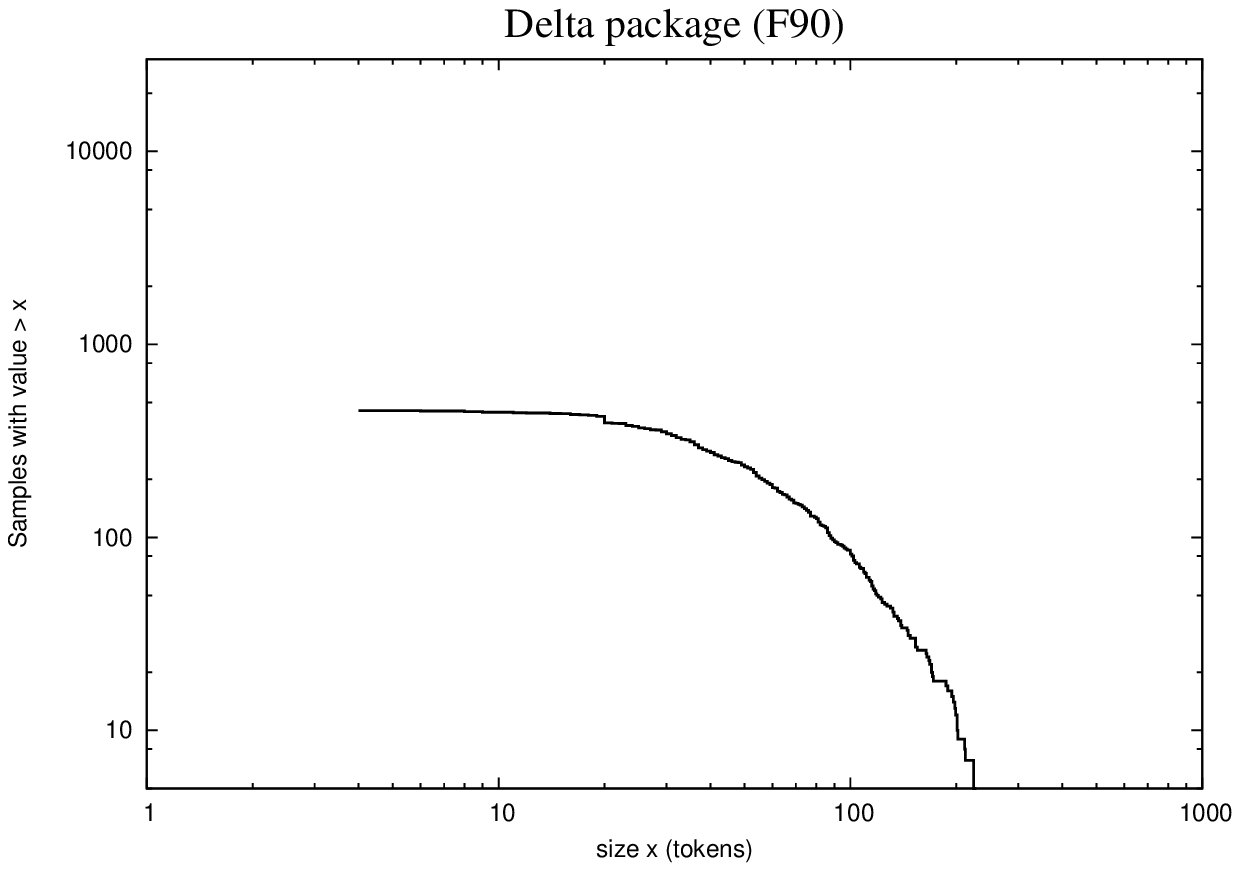,width=0.5\linewidth,clip=} &
\epsfig{file=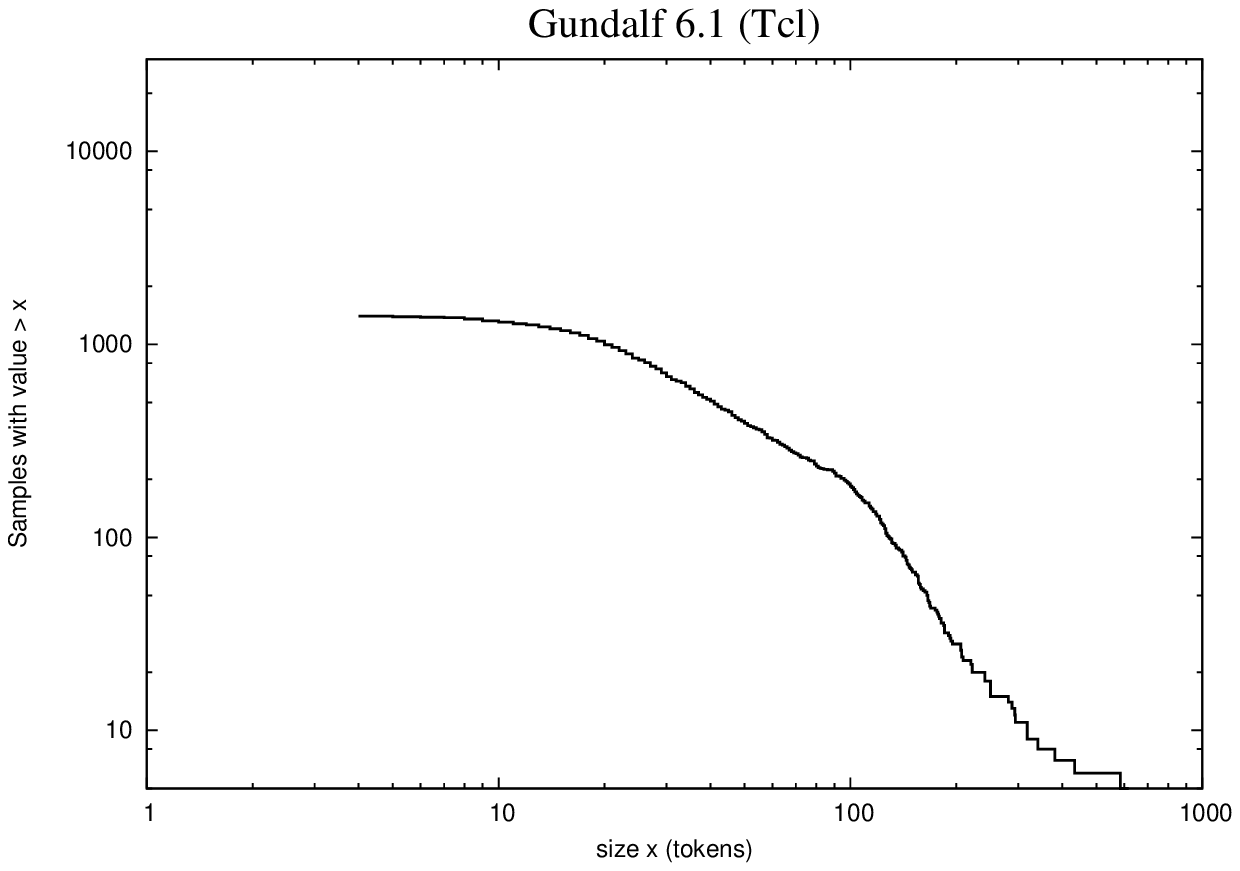,width=0.5\linewidth,clip=} \\
\end{tabular}
\caption{Various Ada, Fortran 77 and 90, C++, Java and Tcl-Tk applications in the range 40,000 - 4,500,000 SLOC.  The y-axis is slightly extended compared with Figures \ref{fig:largec}, \ref{fig:mediumc} and \ref{fig:smallc}.}
\label{fig:variousall}
\end{figure}

In contrast, Figure \ref{fig:variousall} is a collage of various languages of various system sizes as detailed.  Comparing the slight curvature in the tail of the large OO packages in this Figure with the equally large C packages shown in Figure \ref{fig:largec} reveals the effects of the approximation of neglecting nested small methods as described earlier.

For each of Figures \ref{fig:largec}, \ref{fig:mediumc}, \ref{fig:smallc} and \ref{fig:variousall}, the packages decrease in line count from top left to bottom right.  Note that there also some small changes in the y-axis scale.

\subsubsection{Persistence}

Given that the signature of (\ref{eq:pwrlaw}) is visible even in individual packages, it is useful to consider the time of appearance of this behaviour.  Is it present in the first release of a software system or does it emerge as that system is systematically refined during a maintenance cycle ?  The constrained development model described in \cite{Hatton2011a} suggests that the distribution of tokens by programmers under these constraints as they reason about a system unconsciously drives the development at \textit{all} stages.  In other words, it might be expected that this characteristic signature would appear early on in the development process.

In addition, practical considerations suggest that it would be unusual to expect major changes in component size distribution as a system ages on the general grounds that engineers are reluctant to change working systems too much even as they adapt to changing requirements and other normal maintenance activities.

\begin{figure}
\centering
\begin{tabular}{cc}
\epsfig{file=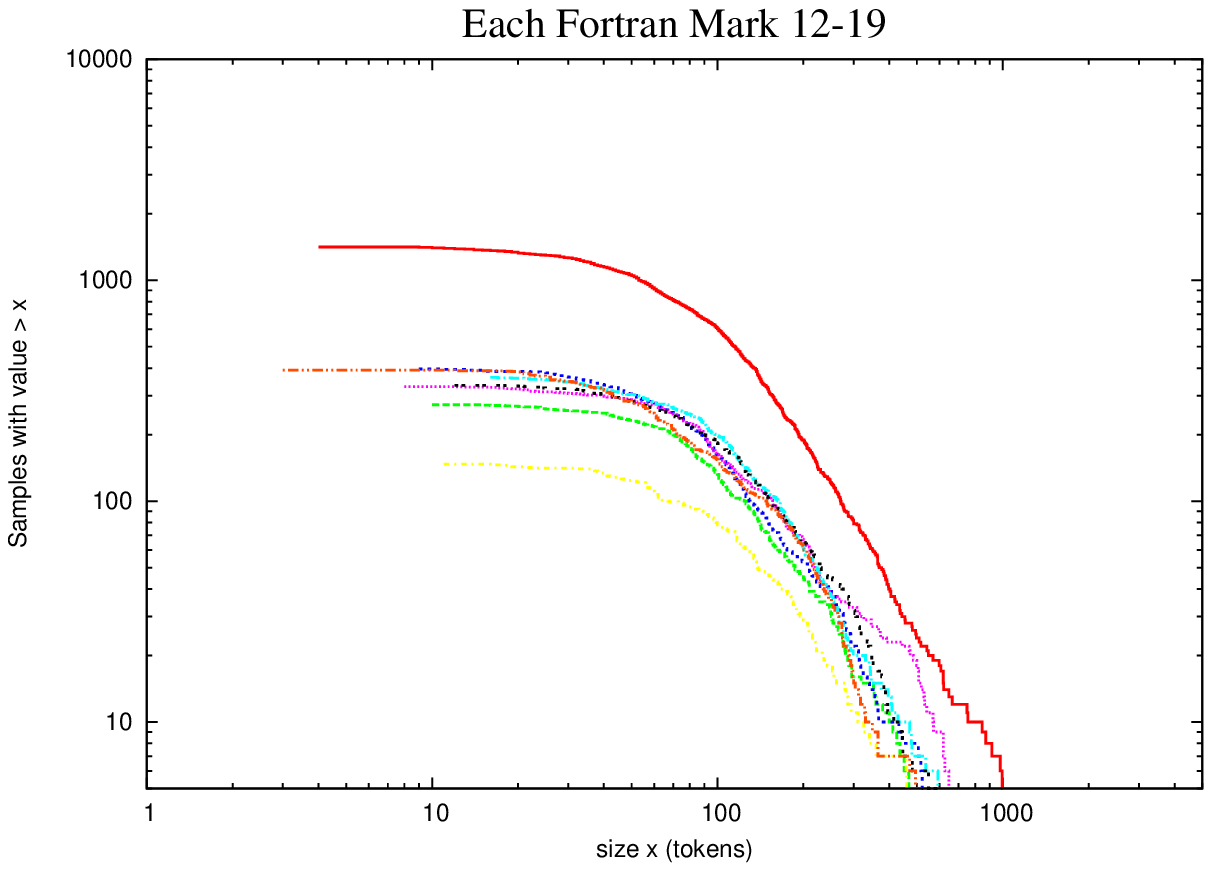,width=0.5\linewidth,clip=} &
\epsfig{file=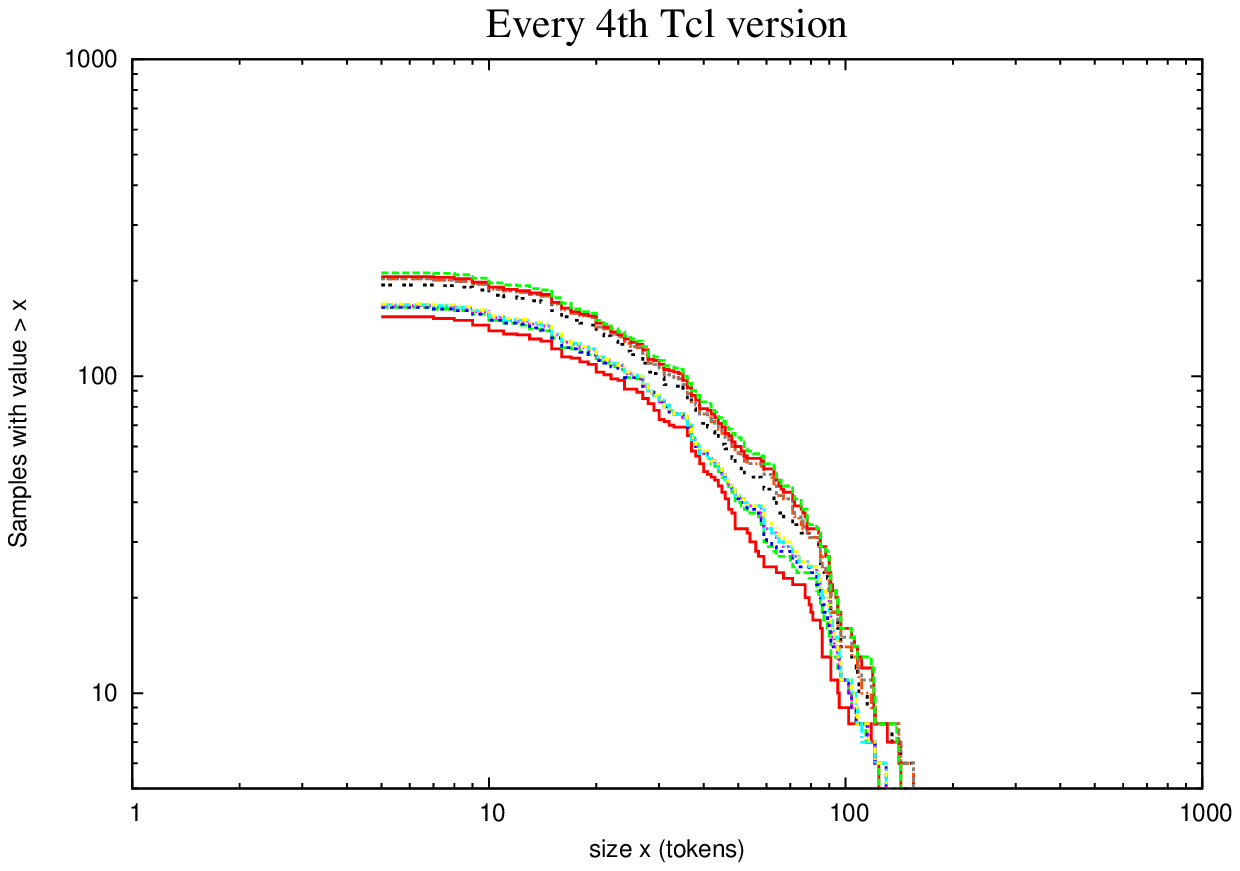,width=0.5\linewidth,clip=} \\
\epsfig{file=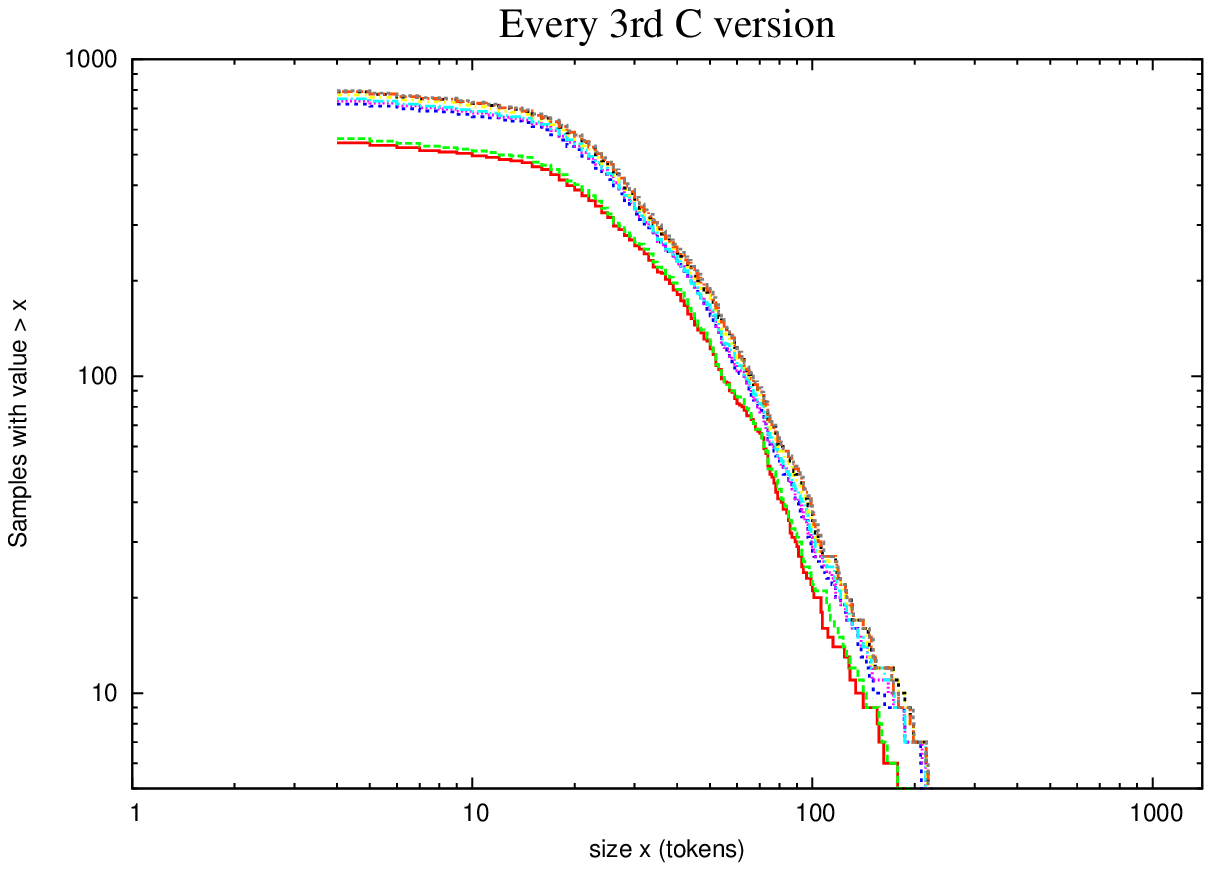,width=0.5\linewidth,clip=}
\end{tabular}
\caption{Distributions of a Fortran, Tcl-Tk and a C system over many releases.}
\label{fig:persistence}
\end{figure}

To address this, the revision history of three very different systems (one Fortran, one in Tcl-Tk and one in C) was analysed.  Two of these were taken from the very first release of 7-8 year life-cycles and one from around half way into its 25 year life-cycle from first release.
\paragraph{}
Relevant parameters of these three packages are shown as Table 1.
\paragraph{}
\begin{tabular}{|p{3cm}|p{2cm}|p{2cm}|p{1cm}|p{1.5cm}|p{1.5cm}|}
\hline Package & Language & Releases & Years & start XLOC & end XLOC \\
\hline
\hline Numerical library & Fortran & 8 & 12 & 90,198 & 266,123 \\ 
\hline Geophysical modelling & Tcl-Tk & 44 & 7 & 6,227 & 11,078 \\ 
\hline Language parser & C & 27 & 8 & 35,851 & 65,270 \\ 
\hline 
\end{tabular}
\paragraph{Table 1: Packages used to show power-law persistence}
\paragraph{}
The left hand upper diagram of Figure \ref{fig:persistence} shows the component size disribution for each official release of a widely used numerical library (the NAG Fortran library) from release 12  through release 19, spanning around 12 years.  The last release analysed, release 19, comprised 3659 components containing altogether almost 270,000 XLOC.  Even though the library almost trebled in size over this period, there is little substantial change in the component size distribution across this time period.  It remains possible that substantial change might have taken place in the releases prior to release 12.  Although the data were not available to confirm this, it would be most unlikely for a scientific subroutine library to change significantly over its life-cycle by the very nature of its functionality.  The solutions of mathematical algorithms hardly vary once implemented.

The right hand diagram of Figure \ref{fig:persistence} shows the component size distributions for every fourth release of a Tcl-Tk system for geophysical modelling as it grew by about a factor of two.

The left hand lower diagram shows every third release of a C system for statically checking C programs as it grew by a factor of two from its first release.

These are very different systems but \textit{the characteristic signature described by (\ref{eq:pwrlaw}) appears to be a persistent property in the evolution of each system, present at first release and preserved during the maintenance cycle, even when that doubles or triples the initial released system size as is the case here}.

\section{Application to genetic systems}
I will now apply the general principle expressed by (\ref{eq:pwrlaw}) to predicting properties of the genome, another discrete token based system, and in particular that of gene length.

\subsection{Genetic background}
The authors of \cite{XuJune2006} surveyed almost all prokaryotic and eukaryotic species whose complete genome sequence data were then available and well annotated. These data included 81 prokaryotes and 19 eukaryotes and regressed the estimate of total coding sequence length against the estimate of the number of genes for each of the two groups of species.  They found that although the average lengths of genes in prokaryotes and eukaryotes are significantly different, \textit{the average lengths of genes are effectively constant within either of the two kingdoms leading to a linear relationship between the total length of the sequence and the number of contributing genes}. They concluded that natural selection has clearly set a strong limitation on gene elongation within the kingdom and that the average gene size adds another distinct characteristic for the discrimination between the two kingdoms of organisms.  Their data can be seen at http://mbe.oxfordjournals.org/content/23/6/1107.full.

Here, I will propose that the reason why the average gene length is highly conserved within a kingdom is inevitable and is related to exactly the same underlying Conservation of Choice or Hartley-Shannon Information so emphatically demonstrated above for software systems.
\section{A variational model for gene length}
First, suppose that genes have length $t_{i}$ bases chosen from a unique alphabet of $a_{i}$ bases.  The complications of defining gene length which include taking account of introns, outrons and so on are avoided by using pre-analysed data as for example in \cite{XuJune2006}.

The key observation is that the alphabet of bases in genetic codes is \textit{fixed} to adenine, cytosine, guanine and thiamine.  In other words, $a_{i} = 4, \forall i$.  In this regard, although the genome is much bigger than any single software system, it is constructed from a much simpler alphabet.  This then implies from (\ref{eq:pwrlaw}) that for such genetic sequences,

\begin{equation}
p_{i} \sim K      \label{eq:pdfconstant}
\end{equation}

where K is a constant.  In other words, Conservation of Information implies that by far the most likely outcome is that \textit{gene lengths are distributed uniformly within whatever kingdom is being considered} and so the average gene length E(L) is constant. Since there are M genes in a total coding length T, we have

\begin{equation}
Constant = E(L) = C \frac{T}{M}
\end{equation} 

where C is some constant depending on the species.  In other words

\begin{equation}
T = k' M
\end{equation} 

where k' is another constant.  This behaviour is precisely that found empirically by \cite{XuJune2006} as shown at http://mbe.oxfordjournals.org/content/23/6/1107.full.

Note that this development says nothing about the kingdom.  It simply says that subject to the total number of bases being constant and the total information being constant in the Hartley-Shannon sense, the overwhelmingly most likely distribution of the lengths of the genes will be uniform leading to the prediction that the average gene length is constant for a kingdom.

It is worthwhile re-iterating the points made by \cite{XuJune2006}.  First of all, they argue ``it is widely accepted that natural selection favours shorter genic coding sequence length for higher transcriptional efficiency, for efficient protein synthesis, and for avoiding accumulation of deleterious mutation. On the other hand however, evolution seems to improve the function of a protein through elongating its coding sequence''.  They go on to say that ``their observations suggest that there is a stringent structural constraint on evolution of gene size on a genomic scale. Furthermore species that have been diverged for more than a few billions of years ago in either prokaryotic (Prochlorococcus marinus) or eukaryotic (Ashbya gossypii) group share a relatively constant mean gene size''.

These comments suggest the existence of a more profound underlying principle at work.  \textit{I propose that this underlying principle is indeed the Conservation of Information.}  In software systems, this has been demonstrated here to be overwhelmingly true whatever language is used, whatever the system does and howsoever the system was built.  By analogy in genetic systems, this underlying clockwork is independent of the nature of evolution.  It merely populates the landscape which evolution traverses with an overwhelming number of places where average gene length is constant for a kingdom.

As a result, it is worth re-iterating Cherry's caution \cite{Cherry1963} against over-emphasizing the relationship between information content and meaning, and by inference, functionality.  The development using information content from equation (\ref{eq:hfunc}) onwards leading to the relationship expressed by equation (\ref{eq:pwrlaw}) is fundamental but it says little if anything about functionality.  Indeed functionality seems irrelevant in the emergence of the properties described by (\ref{eq:pwrlaw}).
\paragraph{}
The proper study of meaning is known as \textit{semiotics}.  In this discipline, rules acting on signs or tokens are split into three categories:-
\begin{itemize}
\item Syntactic rules (rules of syntax; relations between signs)
\item Semantic rules (relations between signs and the things, actions, relationships and qualities known collectively as \textit{designata})
\item Pragmatic rules (relations between signs and their users)
\end{itemize}
The development described here relates only to the first category.

\section{Conclusions}
The paper presents several contributions each supported by real systems data of different provenance.
\begin{itemize}
\item Using variational principles suggested in \cite{HatTSE08}, \cite{Hatton2011a} and using the principle of the Conservation of Information, it is predicted that the probability $p_{i}$ of a component appearing with $t_{i}$ tokens in any software system, whatever its implementation details, obeys the following distribution with respect to the size of its unique alphabet of tokens $a_{i}$,
\begin{equation}
 p_{i} \sim (a_{i})^{- \beta}   \label{eq:lhlaw}
\end{equation}
Overwhelming evidence for this behaviour has been presented derived from some 55.5 million lines of code in six languages with an associated p-value of $< 2.2 . 10^{-16}$.
\item The behaviour exemplified by (\ref{eq:lhlaw}) has been demonstrated to be persistent through the life of single software systems as exemplified by three very disparate systems.
\item The behaviour exemplified by (\ref{eq:lhlaw}) appears with relatively few tokens in software systems. In other words, equilibriation is quite rapid.
\item The underlying principle of Conservation of Information is shown to lead to a prediction that average gene length is constant in a kingdom.  This is supported by independent data. 
\end{itemize}

In summary, the Conservation of Information in discrete token based systems such as all software systems and biological systems such as the genome, appears to play a fundamental role in the development of those systems in a way comparable with the principle of Conservation of Energy in physical systems.

\section{References}

\bibliography{lh_biblio}
\end{document}